\documentclass[preprint,aps,amssymb,showpacs,superscriptaddress,nofootinbib]{revtex4}
\usepackage{graphicx}
\begin{document}

\title{Polymer quantization of CGHS model- I}
\author{Alok Laddha}
\email{alokl@imsc.res.in}
\affiliation{The Institute of Mathematical Sciences,
CIT Campus, Chennai-600 113, INDIA}
\preprint{IMSc/2006/6/16}

\leftline{\today}
\begin{abstract}

We present a polymer(loop) quantization of a two dimensional theory of dilatonic gravity known as the CGHS model. We recast the theory as a parametrized free field theory on a flat 2-dimensional spacetime and quantize the resulting phase space using techniques of loop quantization. The resulting (kinematical) Hilbert space admits a unitary representation of the spacetime diffeomorphism group. We obtain the complete spectrum of the theory using a technique known as group averaging and perform quantization of Dirac observables on the resulting Hilbert space. We argue that the algebra of Dirac observables gets deformed in the quantum theory. Combining the ideas from parametrized field theory with certain relational observables, evolution is defined in the quantum theory in the Heisenberg picture. Finally the dilaton field is quantized on the physical Hilbert space which carries information about quantum geometry.

\end{abstract}

\maketitle

\section{Introduction}

In the past two decades, two dimensional theories of gravity have received quite a bit of attention \cite{moore} as toy models to address questions arising in (four dimensional) quantum gravity. In particular the CGHS model whose action is inspired from the effective target space action of 2-d non-critical string theory constitutes a highly desirable choice, due to its various features such as classical integrability, existence of Black hole spacetimes in its solution space and the presence of Hawking radiation and evaporation at 1-loop level.\\
\hspace*{0.2in} Semi-classical analysis of this model has been carried out by number of authors (\cite{cghs}, \cite{rst}, \cite{s}, \cite{g-s} and ref.therein). By incorporating a large number of conformal scalar fields, Hawking radiation (arising from trace anomaly) and the back reaction take place at the 1-loop level. However during the final stages of the collapse, the semi-classical approximation breaks down signalling a need to incorporate higher order quantum corrections and non-perturbative effects. It is always believed that a non-perturbative quantum theory is required in order to answer questions regarding final fate of singularity, information loss etc. (see however \cite{rst}).\\
\hspace*{0.2in} In the canonical formulation the non-perturbative quantization of CGHS model has been carried out in detail in various papers (\cite{krv}, \cite{jackiw1}, \cite{jackiw2}, \cite{mikovic} and ref.therein). After a rescaling of the metric, the model becomes amenable to Dirac constraint quantization as well as BRST methods. Although the complete spectrum is known in the BRST-approach as well as in the Dirac method (in the so called Heisenberg picture), so far it has not been possible to ask the questions regarding quantum geometry using this spectrum.\\
\hspace*{0.2in} In this paper we begin the analysis of the rescaled-CGHS (KRV) model using the methods of loop quantum gravity (LQG) (\cite{th0}, \cite{r}) more generally known as polymer quantization (\cite{ashtekar1}, \cite{ashtekar2}). More in detail, we derive a quantum theory of dilaton gravity (starting from classical CGHS model) which can be used to understand the near-planckian physics of CGHS model. The aim of this work is two-fold. First, we would eventually like to understand if the methods of loop quantization sheds new light on the structure of quantum geometry close to singularity of the CGHS Black holes. Although we do not answer this question in this paper, we setup a framework where this question can be asked. Secondly as the model offers a greater degree of analytic control than its higher-dimensional avatars, we can study in detail various structures which arise in LQG but have so far remained rather formal. (physical Hilbert space, Dirac observables, relational dynamics)\\
\hspace*{0.2in} We begin by reviewing the classical CGHS model and its canonical formulation in section 2. We recast it as a free parametrized scalar field theory on a fiducial flat space-time \cite{krv}. Parametrized field theories on fixed background have a very rich mathematical and conceptual structure (\cite{hajicek1}, \cite{hajicek2}, \cite{hajicek3}). They are ideal arenas to test various quantization methods which one hopes to use in quantization of General relativity. It is partly our aim to show that by combining the ideas from parametrized field theories and LQG, one obtains a potentially interesting quantum theory of dilaton gravity.\\
\hspace*{0.2in} In section 3 we quantize the phase space at the kinematical level (i.e. prior to solving the constraints). By choosing an appropriate sub-algebra of full Poisson algebra and performing the so called GNS quantization using a positive linear functional (analogous to the Ashtekar-Lewandowski functional used in LQG), we obtain a Hilbert space which carries a unitary representation of the space-time diffeomorphism group of the theory. We use the group averaging method in section 4 to solve the constraints and obtain the complete spectrum (physical Hilbert space) of the theory.\\
\hspace*{0.2in} Parametrized field theories give us a general algorithm to obtain an algebra of Dirac observables (Perennials) of the theory. In section 5 we show how to quantize this algebra on the physical Hilbert space and show how the physical Hilbert space is not a representation space for this algebra (in other words the algebra gets deformed in the quantum theory).\\
\hspace*{0.2in} Using key ideas due to Dittrich \cite{bianca} and Hajicek \cite{hajicek1}, in section 6 we show how to define classical dynamical observables (so called complete observables) for our model. Along with diffeomorphisms of the background spacetime complete observables can be used to define time evolution in the system. We propose two inequivalent quantizations of these observables, and finally define physical dilaton operator on physical Hilbert space which is the primary object in discussions about quantum geometry. We finally conclude with discussion.\\
\pagebreak
\\
\section{Classical theory}

In this section we briefly recall the (rescaled) action of the CGHS model along with the solution of the field equations and the structure of the canonical theory.\\
\hspace*{0.2in} The original CGHS action\footnote{ We choose c=G=1. Thus only basic dimension in the theory is L and [M] = $L^{-1}$. In these units $\hbar$ becomes a dimensionless number.} describing a two dimensional theory of dilatonic gravity is given by,\\
\begin{equation}
S_{CGHS}\; =\; \frac{1}{4}\int d^{2}X\; \sqrt{-g}[\; e^{-2\phi}(\; R[g]\; +\; 4(\nabla \phi)^{2}\; +\; 4\lambda^{2}\; )\; -\; (\nabla f)^{2}\; ].
\end{equation}
\hspace*{0.2in} Here $\phi$ is the dilaton field, g is the spacetime metric (signature (-,+)) and f is a conformally coupled scalar field.\\
Rescaling the metric $g_{\mu\nu}\; =\; e^{2\phi}\; \gamma_{\mu\nu}$ one obtains the KRV action \cite{krv}\\
\begin{equation}
S_{KRV}\; =\; \frac{1}{2}\int d^{2}X\; \sqrt{-\gamma}[\; (\; yR[\gamma]\; +\; 4\lambda^{2}\; )\; -\; \gamma^{\alpha\beta}\nabla_{\alpha}f\nabla_{\beta}f\; )].
\end{equation}
where $y\; =\; e^{-2\phi}$.\\
\hspace*{0.2in} The field equations obtained by varying $S_{KRV}$ can be analyzed in the conformal gauge. The solution is as follows. $\gamma_{\alpha\beta}$ is flat.  The remaining fields can be described most elegantly in terms of null-coordinates
$X^{\pm}\; =\; Z\pm T$ on the flat spacetime.
The scalar field f is simply free field propagating on the flat spacetime\\
\begin{equation}
f(X)\; =\; f_{+}(X^{+})\; +\; f_{-}(X^{-})
\end{equation}
and the dilaton is\\
$y(X)\; =\; \frac{M}{\lambda}\;+\; \lambda^{2}X^{+}X^{-}\; -\frac{1}{2}\int^{X^{+}}d\overline{X}^{+}\int^{\overline{X}^{+}}d\overline{\overline{X}}^{+}\partial_{+}f\partial_{+}f\; -\; \frac{1}{2}\int^{X^{-}}d\overline{X}^{-}\int^{\overline{X}^{-}}d\overline{\overline{X}}^{+}\partial_{-}f\partial_{-}f$.\\
\\
\hspace*{0.2in} Thus the solution space of the original CGHS model, namely $(g_{\mu\nu},f)$ is completely determined in terms of the matter field f. This space contains black hole spacetimes as well. Easiest way to see this is to look at
vacuum solutions. Taking $f(X)\; =\; 0$, one can show that the dilaton is given by,\\
\begin{equation}
y(X)\; =\; \frac{M}{\lambda}\; +\; \lambda^{2}X^{+}X^{-}\;
\end{equation}
and the associated physical metric is,\\
\begin{equation}
g_{\mu\nu}\; =\; \frac{1}{\lambda^{2}X^{+}X^{-}\; + \frac{M}{\lambda}}\gamma_{\mu\nu}
\end{equation}
which correspond to black holes of mass M in 2 dimensions.(M=0 is the linear dilaton vacuum). The singularity occurs where $y(X)\; =\; 0$. One can obtain more generic black hole spacetimes by sending in left-moving matter pulses from past null infinity. In all these cases locus of singularity is defined by $y(X)\; =\; 0$.\footnote{There is an important difference between the CGHS and KRV action at the semi-classical level. In the path integral quantization, Hawking radiation is encoded in a one loop term obtained by integrating out the matter field. This term is known as the Polyakov-Liouville term and is zero if one uses the flat metric $\gamma$ (naturally appearing in the KRV action) to define the measure for the matter field. It is however non-zero if one uses the physical metric g (which appears in the CGHS action). Whence it is often claimed that the theory defined by KRV action does not contain Hawking radiation.\cite{s}}\cite{giddings}

\subsection{Canonical description}

The reason for using the rescaled-KRV action rather than the original (and perhaps more interesting) CGHS action is the following. One can perform a canonical transformation on the canonical co-ordinates of the KRV phase space and obtain a parametrized free field theory on flat background. This will be our starting point for quantization. The details of this canonical transformation (also known as Kuchar decomposition \cite{hajicek4}) are given in \cite{krv}, here we only summarize the main results.\footnote{It is interesting to note that even the phase space of the CGHS action can be mapped onto a parametrized scalar field theory on Kruskal spacetime. However the canonical transformation are singular in a portion of phase space. \cite{m1}}\\
\hspace*{0.2in} The KRV spacetime action can be cast into canonical form by using an arbitrary foliation $X^{\alpha}\; =\; X^{\alpha}(x,t)$ of spacetime by (t=const) spacelike hypersurfaces.\\
\begin{equation}
S_{KRV}\; =\; \int dt\int_{-\infty}^{\infty} dx\; (\pi_{y}\dot{y}\; +\; \pi_{\sigma}\dot{\sigma}\; +\; \pi_{f}\dot{f}\; -NH\; -N^{1}H_{1})
\end{equation}
where $(y(x),\sigma(x),f(x))$ are the pull back of the dilaton, spacetime metric $\gamma_{\mu\nu}$ and the scalar field onto the hypersurface $\Sigma$ respectively and $\pi_{y},\pi_{\sigma},\pi_{f}$ are their conjugate momenta. $(N,N^{1})$ are the usual lapse and shift functions and $H,H_{1}$ are Hamiltonian and momentum constraints respectively and are constrained to vanish. A series of non-local canonical transformations maps the above action into that of a parametrized free field theory of flat background \cite{krv},\\
\begin{equation}\label{eq:zero}
\begin{array}{ll}
S[X^{\pm},\Pi^{\pm},f,\pi_{f},N,N^{1},p,m_{R}]\; =\; \int dt\int_{-\infty}^{\infty}dx (\Pi_{+}\dot{X}^{+}\; +\; \Pi_{-}\dot{X}^{-}\; +\; \pi_{f}\dot{f}\; -\; N\tilde{\tilde{H}}\; -\; N^{1}\tilde{\tilde{H}}_{1})\\
\vspace*{0.2in}
\\
\hspace*{5.0in} +\; \int dt\; p(t)\; \dot{m}_{R}(t)
\end{array}
\end{equation}
where $X^{\pm}(x)$ are the embedding variables and correspond to the light-cone coordinates on the Minkowski spacetime, $\Pi_{\pm}$ are conjugate momenta and the Hamiltonian constraint has been rescaled so as to have the same density weight as the momentum constraint.\footnote{Here our notation is $\tilde{f}$ means f is a density of weight 1.} The boundary term $\int p\dot{m}_{R}$ arises due to asymptotic conditions (Note that there are 2 boundaries in the problem, left and right infinity but only 1 boundary term in the action) on the initial data. $m_{R}$ is the right mass of spacetime and it is conjugate momentum p is the difference between the parametrized time and proper time at right infinity when the parametrized time at left infinity is chosen to agree with the proper time.\footnote{As the physical metric $g_{\mu\nu}$ is asymptotically flat, it has a asymptotic stationary killing field. proper time is the time measured by clock along the orbit is of this killing field. Parametrized time is the time defined by asymptotic value of the lapse function. For more details see \cite{k1}}. This action is the canonical action for a parametrized massless scalar field theory on flat spacetime.\\
\hspace*{0.2in} The 2 constraints can be combined to form two Virasoro constraints $\tilde{\tilde{H}}^{\pm}\; =\; \frac{1}{2}(\tilde{\tilde{H}}\; \pm\; \tilde{\tilde{H}}_{1})$. These two Virasoro constraints mutually commute with each other. Thus the constraint algebra can be written as a direct sum of two Lie algebras each of which generates Diff($\bold{R}$).\\
We now proceed to the quantum theory.\\

\section{Quantum theory}

\hspace*{0.2in} In this section we quantize the classical theory using the techniques of polymer quantization. To make this section self contained we recall the basic idea behind polymer quantization in a manner that is coherent with the recent work on kinematical quantization in LQG. (\cite{th1}, \cite{th2}, ref.therein)\\

The basic idea behind polymer quantization can be summarized as follows.\\

1. Choose an appropriate sub-algebra of the full Poisson algebra, which does not use any background metric.\\
2. Define corresponding quantum algebra. This is an abstract *-algebra with an associative multiplication and in which the Poisson relations are represented as Lie-relations. As our quantum algebra does not use any background metric, it turns out that diffeomorphisms act as group of outer automorphisms on this algebra.\\
3. Choose a diffeomorphism invariant positive linear functional to construct a (cyclic) representation of this algebra, using GNS quantization.\\
The resulting Hilbert space then carries a unitary and anomaly-free representation of diffeomorphism-group of the theory.\\

These steps are rather generic to canonical quantization of field theories. The difference between polymer and Fock quantization lies in choice of the sub algebra which we are interested in quantizing and more importantly in the choice of the GNS-functional. \\
We now carry out these steps in detail for our model.

\subsection{Embedding sector}

The first step towards canonical quantization is a suitable choice of quantum algebra. Let us first describe our choice of quantum algebra for the embedding sector. Recall that $\Pi^{\pm}$ are scalar densities of weight +1 (equivalently 1-forms in 1-dim.) and $X^{\pm}$ are scalars (equivalently densitized vector fields).\\

\hspace*{0.2in} Consider a graph $\gamma$ in the spatial slice $\Sigma$ as a collection of finite number of edges and vertices. Define a cylindrical function for both the right moving(+) and left-moving(-) embedding sectors as\\
\begin{equation}
f^{\pm}_{\gamma}\; =\; \Pi_{e\in E(\gamma)}\; exp(ik_{e}^{\pm}\; \int_{e}\Pi^{\pm})
\end{equation}
where $k_{e}^{\pm}\in \bold{R}$.\\

\hspace*{0.2in} Define abelian *-algebras $Cyl^{\pm}\; =\; \cup_{\gamma\in \Gamma}\; Cyl_{\gamma}^{\pm}$. Let Vec denote the complexified Lie algebra of vector fields $X^{\pm}(x)$ which are maps $Cyl^{\pm}$ $\rightarrow$ $Cyl^{\pm}$, (via Poisson brackets) that satisfy Leibniz rule and annihilate constants.\\

For both, the left-moving as well as right-moving sectors, consider the Lie-* algebra V defined by,\\
\begin{equation}
[(f,X(x)),(f',X'(x'))]\; =\; (\lbrace{X(x),f'\rbrace}\; -\; \lbrace{X'(x'),f\rbrace}\; ,\; 0)
\end{equation}
where $(f,f')$ are in Cyl and (X,$X'$) are vector fields. *-operation is just complex conjugation. (Conjugation of vector fields is defined by $X(x)^{*}f\ :=\; (X(x)f)^{*}$.)\\
\hspace*{0.2in} We now define the quantum algebras for the embedding sector. Our derivation mimics the derivation of quantum algebra for LQG given in \cite{th2}.\\
\hspace*{0.2in} Let us denote the pair $(f^{\pm},X^{\pm}(x))$ by a symbol $a^{\pm}$. Consider the *-algebra of finite linear combinations of finite sequences of the form $(a^{\pm}_{1},...,a^{\pm}_{n})$ with an associative product,\\
\begin{equation}
(a^{\pm}_{1},...,a^{\pm}_{n})\cdot(a^{\pm}_{n+1},...,a^{\pm}_{m})\; =\; (a^{\pm}_{1},...,a^{\pm}_{m})
\end{equation}
and an involution,\\
\begin{equation}
(a^{\pm}_{1},...,a^{\pm}_{n})^{*}\; =\; (a^{\pm*}_{n},...,a^{\pm*}_{1})
\end{equation}
Divide this algebra by a two sided ideal defined by elements of the form:
\begin{equation}
\begin{array}{ll}
(ka^{\pm})\; -\; k(a^{\pm})\\
\vspace*{0.1in}
(a^{\pm}_{1}\; +\; a^{\pm}_{2})\; -\; (a^{\pm}_{1})\; -\; (a^{\pm}_{2})
\end{array}
\end{equation}
\hspace*{0.2in} The resulting algebras (for both $\pm$ sectors) are nothing but the free tensor algebras generated by $a^{\pm}$. The algebras $\mathcal{U}_{E}^{\pm}$  that we will quantize are defined as the free tensor algebras defined above modulo the 2-sided ideal generated by elements of the form $a_{1}^{\pm}\otimes a_{2}^{\pm}\; -\; a_{2}^{\pm}\otimes a_{1}^{\pm}\; -\; [a_{1}^{\pm},a_{2}^{\pm}]$.\footnote{$\mathcal{U}_{E}^{\pm}$ is nothing but the universal enveloping algebra of the Lie algebra $V^{\pm}$.}\\

\hspace*{0.2in} So finally the algebra that we choose for quantization is $\mathcal{U}_{E}\; =\; \mathcal{U}_{E}^{+}\otimes\mathcal{U}_{E}^{-}$.\\
\hspace*{0.2in} The group generated by the two Virasoro constraints which is a direct product of two copies of Diff($\bold{R}$) has a natural representation as a group of outer automorphisms on $\mathcal{U}_{E}$. Abusing the standard nomenclature we refer to this group as Virasoro group.\\
\begin{equation}
\alpha_{\phi}^{\pm}(f_{\gamma}^{\pm})\; =\; f^{\pm}_{\phi^{-1}(\gamma)}
\end{equation}
\\
\begin{equation}
\alpha_{\phi}^{\pm}(X^{\pm}(x))\; =\; X^{\pm}(\phi^{-1}(x))
\end{equation}
\hspace*{0.2in} The representation of $\mathcal{U}_{E}$ should be such that the outer automorphisms of $\mathcal{U}_{E}$ are represented via unitary operators as inner automorphisms. i.e.\\
\begin{equation}
U_{\pi}(\phi^{\pm})\pi(a^{\pm})U_{\pi}(\phi^{\pm})^{-1}\; =\; \pi(\alpha_{\phi^{\pm}}(a^{\pm}))\;\; \forall a\in \mathcal{U}_{E}
\end{equation}
\\
\hspace*{0.2in} By GNS-quantization of the C* sub-algebra generated by $Cyl^{\pm}$ \cite{th1} one obtains a diffeomorphism-invariant irreducible representation of $\mathcal{U}_{E}$. The representation is given by $\mathcal{H}_{E}^{\pm}\; =\; L^{2}(\overline{\mathcal{\pi}}^{\pm}\; ,\; d\mu_{0})$ where $\overline{\mathcal{\pi}}^{\pm}$ is the spectrum of C* sub-algebra generated by $Cyl^{\pm}$ and $d\mu_{0}$ is a regular Borel probability measure given by\\
\begin{equation}
\int f_{\gamma}^{\pm}\; d\mu_{0}\; =\; \omega_{0}^{\pm}(f_{\gamma}^{\pm})
\end{equation}
where $\omega_{0}^{\pm}$ is a positive linear functional which is motivated by the A-L positive linear functional of LQG.\\
\begin{equation}
\omega_{0}^{\pm}(f_{\gamma}^{\pm})\; =\; \delta_{\gamma,0}
\end{equation}
\hspace*{0.2in} On $\mathcal{H}_{E}^{\pm}$ cylindrical functions act as multiplication operators and one can show that the embedding variables act as derivations,\\
\begin{equation}
\begin{array}{lll}
\hat{X}^{\pm}(x)f_{\gamma}^{\pm}(\Pi^{\pm})\; & = & (-i\hbar)\; ik_{e}\; f_{\gamma}^{\pm}(\Pi^{\pm})\; \; if\; x\in e\\
\vspace*{0.1in}
& = & (-i\hbar)\; i\frac{(k_{e}+k_{e'})}{2}\; f_{\gamma}^{\pm}(\Pi^{\pm})\; \; if\; x\in e\cap e'\\
\vspace*{0.1in}
& = & \; 0\; otherwise
\end{array}
\end{equation}
The Virasoro group acts unitarily on $\mathcal{H}_{E}^{\pm}$  as\\
\\
\begin{equation}
\begin{array}{lll}
\hat{U}^{\pm}(\phi^{\pm})\; f^{\pm}_{\gamma}(\pi^{\pm}) =\; f^{\pm}_{(\phi^{\pm})^{-1}\gamma}(\pi^{\pm})\\
\vspace*{0.1in}
\hat{U}^{\pm}(\phi^{\pm})\; \hat{f}^{\pm}_{\gamma}\; \hat{U}^{\pm}(\phi^{\pm})\; =\; \hat{f}^{\pm}_{(\phi^{\pm})^{-1}\gamma}\\
\vspace*{0.1in}
\hat{U}^{\pm}(\phi^{\pm})\; \hat{X}^{\pm}(x)\; \hat{U}^{\pm}(\phi^{\pm})\; =\; \hat{X}^{\pm}(\phi^{\pm}(x))
\end{array}
\end{equation}
The complete embedding Hilbert space is of course given by $\mathcal{H}_{E}\; =\; \mathcal{H}_{E}^{+}\otimes \mathcal{H}_{E}^{-}$.\\
\\
\subsection{Matter sector}

\hspace*{0.2in} Now we consider the kinametical quantization of the matter sector. The quantization given here is unitarily inequivalent to the Bohr quantization of scalar field but it is the same quantization that is used by Thiemann to quantize the Bosonic string. For more details we refer the reader to \cite{th3}.\\

\hspace*{0.2in} Once again the choice of quantum algebra will be motivated by the fact that we want the Virasoro group to act as group of outer automorphisms on this algebra.\\
Following observations help us make such a choice.\\

\hspace*{0.2in} Consider the canonical transformation $(\pi_{f},f)\; \rightarrow\; (Y^{\pm}\; =\; \pi_{f}\ \pm\ f')$. $Y^{\pm}$ satisfy the Poisson bracket relations,\\

\begin{equation}\label{eq:1}
\begin{array}{lll}
\lbrace{\; Y^{\pm}(x)\; ,\; Y^{\pm}(x')\; \rbrace}\; =\; \mp(\; \partial_{x'}\delta(x',x)\; -\; \partial_{x}\delta(x,x')\; )\\
\vspace*{0.1in}
\lbrace{\; Y^{\pm}(x)\; ,\; Y^{\mp}(x')\; \rbrace}\; =\; 0.
\end{array}
\end{equation}

In terms of these variables the Virasoro constraints are given by,\\
\begin{equation}
\begin{array}{lll}
H^{+}(x)\; =\; \Pi_{+}X^{+'}\; +\; \frac{1}{4}(\pi_{f}\; +\; f')^{2}\\
\vspace*{0.1in}
H^{-}(x)\; =\; \Pi_{-}X^{-'}\; -\; \frac{1}{4}(\pi_{f}\; -\; f')^{2}
\end{array}
\end{equation}
Whence one can see that under the Lie-derivative along the Hamiltonian vector field of the constraints,\\
\begin{equation}
\begin{array}{lll}
\mathcal{L}_{H^{\pm}[N_{\pm}]}\; Y^{\pm}(x)\; =\; (N_{\pm}\; Y^{\pm})'(x)\\
\vspace*{0.1in}
\mathcal{L}_{H^{\pm}[N_{\pm}]}\; Y^{\mp}(x)\; =\; 0.
\end{array}
\end{equation}
Thus it is clear that the 2 generators of the Virasoro algebra $H^{\pm}$ act as generators of spatial diffeomorphisms on $Y^{\pm}$. These considerations motivate the following.\\

\hspace*{0.2in} Once again let $\Gamma$ be the set of all graphs $\gamma$ embedded in $\Sigma$ consisting of finite number of edges and vertices. We start by defining momentum network (similar to spin-network in LQG) as a pair $(\gamma,\; \vec{l}(\gamma)\; :=\; (l_{e_{1}},...,l_{e_{N}}))$ where $l_{e}$ are real numbers. A momentum network operator for both the right and left moving sectors is defined as,\\
\begin{equation}
W^{\pm}(s)\; :=\; exp(i\; [\; \sum_{e\in E(\gamma)}l_{e}^{\pm}\; \int_{e}\; Y^{\pm}]\; )
\end{equation}
\hspace*{0.2in} The Weyl relations obeyed by $W^{\pm}(s)$ can be easily derived from (\ref{eq:1}) (using BHC formula),
\begin{equation}\label{eq:2}
\begin{array}{lll}
W^{\pm}(s_{1})W^{\pm}(s_{2})\; =\; e^{\mp\frac{i\hbar}{2}[\alpha(s_{1},s_{2})]}W^{\pm}(s_{1}\; +\; s_{2})\\
\vspace*{0.1in}
W^{\pm}(s)^{*}\; =\; W^{\pm}(-s)
\end{array}
\end{equation}
where\\
\begin{equation}
\alpha(s_{1},s_{2})\; =\; \sum_{e_{1}\in \gamma(s_{1})}\sum_{e_{2}\in \gamma(s_{2})}l^{e_{1}}(s_{1})l^{e_{2}}(s_{2})\alpha(e_{1},e_{2})
\end{equation}
\\
with $\alpha(e_{1},e_{2})\; =\; [\kappa_{e_{1}}]_{\partial e_{2}}\; -\; [\kappa_{e_{2}}]_{\partial e_{1}}$. Here $\kappa_{e}$ is the characteristic function of e. In (\ref{eq:2}) notation $(s_{1}\; +\; s_{2})$ means we decompose all edges $e_{1}$ and $e_{2}$ in their maximal mutually non-overlapping segments and assign $l^{e_{1}}+l^{e_{2}}$ to $e_{1}\cap e_{2}$, $l^{e_{1}}$ to $e_{1}\; -\; \gamma(s_{2})$ and $l^{e_{2}}$ to $e_{2}\; -\; \gamma(s_{1})$ respectively.\\

\hspace*{0.2in} Now we define the algebra that we will be interested in quantizing. Consider an associative algebra generated by formal finite linear combinations of formal sequences of the form $(W^{\pm}_{s_{1}},...,W^{\pm}_{s_{n}})$ with associative multiplication given by,

\begin{equation}
(W^{\pm}_{s_{1}},...,W^{\pm}_{s_{n}})\cdot (W^{\pm}_{s_{n+1}},...,W^{\pm}_{s_{m}}):=\; (W^{\pm}_{s_{1}},...,W^{\pm}_{s_{m}})
\end{equation}
We give this algebra tensor product structure by moding out 2-sided ideals generated by elements of the form,\\
\begin{equation}
\begin{array}{lll}
(\alpha\; W^{\pm}(s))\; -\; \alpha(W^{\pm}(s))\; \alpha \in \mathbb{C}\\
\vspace*{0.1in}
(W^{\pm}(s_{1})\; +\; W^{\pm}(s_{2}))\; -\; (W^{\pm}(s_{1}))\; -\; (W^{\pm}(s_{2}))
\end{array}
\end{equation}
\hspace*{0.2in} We refer to this tensor algebra as $Cyl^{\pm}_{M}$. The *-algebra that we will quantize is $Cyl^{\pm}_{M}$ modulo the 2 sided ideal implied by (\ref{eq:2}).We denote this algebra as $\mathcal{U}^{\pm}_{M}$. Finally the full algebra for both sectors is given by $\mathcal{U}_{M}\; =\; \mathcal{U}^{+}_{M}\otimes \mathcal{U}^{-}_{M}$.

\hspace*{0.2in} As emphasized earlier, the reason for choosing this particular algebra for quantization is its covariance properties under action of the Virasoro group.\\

For all momentum-network functions $W^{\pm}(s)$,\\
\begin{equation}
\alpha_{\phi}^{\pm}(W^{\pm}(s))\; =\; W^{\pm}(\phi(s))
\end{equation}
where $\phi(s):= (\phi^{-1}(\gamma)\; ,\; \vec{l}(\gamma))$.\\
\hspace*{0.2in}Now just like for the embedding sector we perform a GNS quantization of $\mathcal{U}_{M}$ using a Virasoro-invariant positive linear functional,\\
\\
\begin{equation}
\omega_{\pm}(W^{\pm}(s))\; =\; \delta_{s,0}
\end{equation}
where 0 in $\delta_{s,0}$ stands for a graph with zero edges and an empty label set.\\

\hspace*{0.2in} This functional is clearly motivated by the Ashtekar-Lewandowski functional used in LQG. It can be easily shown to be virasoro-invariant. The resulting Hilbert space for both ($\pm$) sectors is given by Cauchy completion of $\mathcal{U}_{M}^{\pm}$. and the representation is,\\
\begin{equation}
\hat{W}^{\pm}(s_{1})\; W^{\pm}(s_{2})(Y^{\pm})\; =\; W^{\pm}(s_{1})(Y^{\pm})W^{\pm}(s_{2})(Y^{\pm})\; =\; e^{\mp\frac{i\hbar}{2}[\alpha(s_{1},s_{2})]}W^{\pm}(s_{1}\; +\; s_{2})(Y^{\pm})
\end{equation}
\hspace*{0.2in} As $\omega_{\pm}$ is Virasoro invariant, it implies that the Virasoro group acts unitarily and anomaly-freely on $\mathcal{H}^{\pm}_{M}$\\
\begin{equation}
\begin{array}{lll}
\hat{U}^{\pm}(\phi^{\pm})\; W^{\pm}(s)(Y^{\pm}) =\; W^{\pm}( (\phi^{\pm})s)(Y^{\pm})\\
\vspace*{0.1in}
\hat{U}^{\pm}(\phi^{\pm})\; \hat{W}^{\pm}(s)\; \hat{U}^{{\pm}^{-1}}(\phi^{\pm})\; =\; \hat{W}^{\pm}(\phi^{\pm}s)
\end{array}
\end{equation}
\hspace*{0.2in} Finally the Matter Hilbert space is $\mathcal{H}_{M}\; =\; \mathcal{H}^{+}_{M}\otimes \mathcal{H}^{-}_{M}$.\footnote{Note that this Hilbert space is not of the form $L^{2}(\overline{Y^{\pm}},d\mu)$. It is (the completion of) algebra itself considered as a vector space with inner product defined by the GNS state.}
\\
\subsection{Quantizing the asymptotic degrees of freedom}

\hspace*{0.2in} The final component of kinematical Hilbert space is the Schroedinger representation of the boundary data $(m,\; p)$. As the Virasoro group has trivial action on the corresponding Heisenberg algebra, we do not need to loop quantize these asymptotic degrees of freedom. Thus we choose for the boundary Hilbert space $L^{2}(\mathbf{R}, dm)$\footnote{More precisely we need to perform quantization on a half-line in order to restrict ourselves to $m \geq 0$ configurations.}
 with,\\
\begin{equation}
\begin{array}{lll}
\hat{m}_{R}\, \Psi(m)\; =\; m_{R}\, \Psi(m)\\
\vspace*{0.1in}
\hat{p}\, \Psi(m)\; =\; -i\frac{\partial}{\partial m}\, \Psi(m)
\end{array}
\end{equation}
\hspace*{0.2in} This finishes the construction of kinematical Hilbert space of the quantum theory. We rewrite the final Hilbert space as tensor product of 3 Hilbert spaces,\\
\begin{equation}
\begin{array}{lll}
\mathcal{H}_{kin}\; & = & \mathcal{H}_{M}\otimes \mathcal{H}_{E}\; \otimes \mathcal{H}_{m} =\; (\mathcal{H}^{+}_{M}\otimes \mathcal{H}^{-}_{M})\otimes (\mathcal{H}^{+}_{E}\otimes \mathcal{H}^{-}_{E})\otimes \mathcal{H}_{m}\\
\vspace*{0.1in}
& = & (\mathcal{H}^{+}_{M}\otimes \mathcal{H}^{+}_{E})\otimes (\mathcal{H}^{-}_{M}\otimes \mathcal{H}^{-}_{E})\otimes \mathcal{H}_{m}.
\end{array}
\end{equation}
\hspace*{0.2in} one for the left moving(-) sector ($\mathcal{H}^{+}$), one for the right moving(+) sector ($\mathcal{H}^{-}$) and one for the asymptotic sector. The Virasoro group acts unitarily on the Hilbert space as 2 mutually commuting copies of spatial diffeomorphisms.\\
We define the basis for $\mathcal{H}$ as follows.\\
\hspace*{0.2in} Consider a graph $\gamma$ with a set of pair of real numbers $((k_{1}^{\pm},l_{1}^{\pm}),...,(k_{N}^{\pm},l_{N}^{\pm}))$ where in outermost pairs ($(k_{1}^{\pm},l_{1}^{\pm})$ and $(k_{N}^{\pm},l_{N}^{\pm}))$) either $k_{i}$ or $l_{i}$ can be zero but not both. in the interior edges $(e_{2},...,e_{N-1})$ we even allow both the charges (k,l) to be zero.

\hspace*{0.2in} We call the pair $(\gamma, ((k_{1}^{\pm},l_{1}^{\pm}),...,(k_{N}^{\pm},l_{N}^{\pm})))$ charge-network (in analogy with spin-networks in LQG) and denote it by s. The state associated with s is given by,\\

\begin{equation}
f_{s}^{\pm}\; =\; e^{\sum_{e\in E(\gamma)}ik^{\pm}_{e}\int_{e} \pi^{\pm}} e^{i\sum_{e\in E(\gamma)}l^{\pm}_{e}\int_{e} Y^{\pm}}
\end{equation}

\section{Physical Hilbert space}

\hspace*{0.2in} The motivation behind choosing a particular quantum algebra and a peculiar GNS functional  has been the unitary and anomaly-free representation of the Virasoro group on the Hilbert space. This is analogous to the representation of spatial diffeomorphisms in LQG whence we use the same method that is used there to solve the diffeomorphism constraint to solve the Virasoro constraints. It is more commonly known as group averaging \cite{marolf}. The idea is to construct a rigging map $\eta$ from a dense subspace $\Phi_{kin}$ of $\mathcal{H}_{kin}$ to its algebraic dual $\Phi^{*}_{kin}$ such that the image of the map are solutions to the Virasoro constraints in the following sense.\\

\begin{equation}
\Psi(\hat{U}(\phi^{\pm})f_{s}^{\pm})\; =\; \Psi(f_{s}^{\pm})\; \; \forall \phi\in Diff(\bold{R})
\end{equation}
\hspace*{0.2in} The rigging map is defined as follows. Given a charge network s, define $\lbrace{[s]\; =\; \phi\cdot s\vert \phi\in Diff(\bold{R})\rbrace}$.\\
Then the rigging map (which is tied to the charge-network basis) is given by,\\
\begin{equation}
\eta(f^{+}_{s}\otimes f^{-}_{s'})\; :=\; (\eta^{+}\otimes\eta^{-})(f^{+}_{s}\otimes f^{-}_{s'})\; =\; \sum_{\phi\in Diff_{[s]}(\bold{R})}\; <\; f^{+}_{s}\hat{U}^{+}(\phi)\; ,\; >\otimes \sum_{\phi'\in Diff_{[s']}(\bold{R})}\; <\; f^{-}_{s'}\hat{U}^{-}(\phi')\; ,\; >
\end{equation}
where the sum is over all the diffeomorphisms $\phi$ which take a charge-network $s\; =\; (\gamma, (\vec{k^{\pm}}, \vec{l^{\pm}}))$ to a different charge network $s'\; =\; (\phi^{-1}(\gamma), (\vec{k^{\pm}}, \vec{l^{\pm}}))$.
In higher dimensions (i.e. when dimension of spatial slice is greater than 1) this map does not work, as the orbits ( set of diffeomorphisms), over which we are averaging, can be infinitely different even for two states based on the same graph. This, in turn, poses problem in defining the inner product on the vector space of group averaged states (\cite{ashtekar3} , \cite{th0}).\footnote{In fact the rigging map in LQG is not even derived in full generality to the best knowledge of the author,\\ and is only  defined for the so called strongly diffeo-invariant observables.} However in 1 dimension the above map yields solutions to the Virasoro constraints. As can be clearly seen from the definition of rigging map, the solution space is a tensor product of 2 vector spaces. Inner product on both of them can be defined as,\\
\begin{equation}
<\eta^{\pm}(f^{\pm}_{s})\vert \eta^{\pm}(f^{\pm}_{s'})>\; =\; [\eta^{\pm}(f^{\pm}_{s'})](f^{\pm}_{s})
\end{equation}
\hspace*{0.2in} The physical Hilbert space is thus characterized by diffeomorphism-equivalence class of charge networks which in 1 dimensions can be classified by the following data.\\
1. Number of edges $\vert E(\gamma)\vert\; =\; N$\\
2. Number of vertices $\vert V(\gamma)\vert\; =\; N+1$\\
3.The set $(\theta_{1},...\theta_{N-1})$ which is the set consisting of the degree of differentiability of graph at each vertex except at the two outermost vertices.\\
4.The ordered set $((k_{1}^{\pm},l_{1}^{\pm}),...,(k_{N}^{\pm},l_{N}^{\pm}))$\\
Unlike in higher dimensions there are no continuous moduli.\\
Thus we can write a ket in $\mathcal{H}^{\pm}_{phy}$ as\\
\begin{equation}
\vert \Psi>_{\pm}\; =\; \vert N,\; N+1,\; (\theta_{1},...\theta_{N-1}),\; (k_{1}^{\pm},l_{1}^{\pm}),...,(k_{N}^{\pm},l_{N}^{\pm})>
\end{equation}
\hspace*{0.2in} Finally as the Virasoro group acts trivially on $\mathcal{H}_{m}$ it remains unchanged under group averaging, whence complete $\mathcal{H}_{phy}\; =\; \mathcal{H}^{+}_{phy}\otimes \mathcal{H}^{-}_{phy}\otimes \mathcal{H}_{m}$.\\
\hspace*{0.2in} At this point we would like to comment on the anomaly-freeness of our representation. In the Fock space quantization of the model (\cite{krv}, \cite{jackiw2}), one obtains a Virasoro anomaly in the constraint algebra due to the Schwinger term in the commutator of the energy momentum tensor for the matter field. In \cite{krv}, the anomaly is removed by modifying the Embedding sector of the theory where as when one uses BRST methods to quantize the model, anomaly is removed by adding background charges(enhancing the central charge) and ghost fields(which define so called bc-CFT). Our choice of Poisson sub-algebra coupled with a unusual choice of GNS functional results in a discontinuous(but anomaly free) representation  of the Virasoro group. Here it is important to note that even in Fock space one can normal order the constraints with respect to so called squeezed vacuum state \cite{jackiw2} such that the central charge is zero. However these states have peculiar properties like the action of finite gauge transformations is ill-defined on them. Contrast this with our representation where infinitesimal gauge transformations are ill-defined.\\

\section{Complete set of Dirac observables}

\hspace*{0.2in} By group averaging the Virasoro constraints we have obtained a physical Hilbert space $\mathcal{H}_{phy}\ =\ \mathcal{H}_{phy}^{+}\otimes\ \mathcal{H}_{phy}^{-}$. Now we encounter (what one always encounters at some stage in canonical quantization of diffeo-invariant theories) problem of time. There is apriori no dynamics on the physical Hilbert space. In order to ask the dynamical questions about for e.g. singularity resulting from the collapse of scalar field in quantum theory, some notion of dynamics should be defined on $\mathcal{H}_{phy}$. We do this by employing ideas due to \cite{bianca}, \cite{hajicek1} which goes back to the old idea of evolving constants of motion by \cite{rovelli}.\\
\hspace*{0.2in} However first we show how to define a complete set of Dirac observables(Perennials) for our model and how to represent them as well-defined operators on $\mathcal{H}_{phy}$. For the classical theory these perennials have been known for a long time ( \cite{krv} , \cite{torre} ) and are analogous to the DDF observables of bosonic string theory \cite{urs}.\\
\hspace*{0.2in} The basic idea behind constructing Dirac observables in parametrized field theory is fairly simple.(This is a general algorithm for defining Dirac observables in parametrized field theories and is also known as Kuchar decomposition \cite{hajicek4}.) Given the phase space of the theory co-ordinatized by $(X^{\pm},\ \Pi^{\pm},\ f,\ \pi_{f})$, one can perform a canonical transformation to the so-called Heisenberg chart $( X^{\pm},\ \bold{\Pi}^{\pm},\ \bold{f},\ \bold{\pi_{f}})$ \cite{krv} where  $\bold{\Pi}^{\pm}$ are the two Virasoro constraints and $(\bold{f},\ \bold{\pi_{f}})$ are the scalar field data on an initial(fixed) slice. $( X^{\pm},\ \bold{\Pi}^{\pm})$ and $(\bold{f},\ \bold{\pi_{f}})$ form a mutually commuting canonically conjugate pair whence it is clear that $(\bold{f},\ \bold{\pi_{f}})$ are Dirac observables.\\
\hspace*{0.2in} Choosing the initial slice as $(X^{\pm}_{0}(x)\ =\ x)$ we can expand these observables in terms of an orthonormal set of mode functions  $e^{ikx}$,\\
\begin{displaymath}
\bold{f}(x)\ =\ \frac{1}{2\sqrt{\pi}}\int_{-\infty}^{\infty}\frac{dk}{\vert k\vert}e^{ikx}a_{k}\ +\ c.c.
\end{displaymath}
\hspace*{0.2in} It is clear that $(a_{k} , a^{*}_{k})$ are also Dirac observables. It is also clear in the Heisenberg chart that they form a complete set (describe true degrees of freedom of the theory). Now by expressing $a_{k}$ in terms of the original (Schroedinger) canonical chart we will obtain Dirac observables that will be promoted to operators on $\mathcal{H}_{phy}$.\\
\hspace*{0.2in} In order to write $a_{k}\ =\ a_{k}[X^{\pm},\Pi^{\pm},f,\pi_{f}]$ one has to appeal to the spacetime picture of a parametrized field theory propagating on flat background. We just summarize the main results and refer the reader to \cite{torre} for details.\\
The scalar field f(X) in spacetime satisfies,\\
\begin{displaymath}
\Box f(X)\ =\ 0
\end{displaymath}
The solution can be expanded as,\\
\begin{displaymath}
f(X)\ =\ \int \frac{dk}{\vert k\vert} e^{ik\cdot X}a_{k}\ +\ c.c.
\end{displaymath}
$a_{k}$'s can be projected out of the solutions f(X)'s on any hypersurface using the Klein-Gordon inner product.\\
\begin{equation}
a_{k}\; =\; i\int_{\Sigma}\; \sqrt{g}[\; e^{-ik\cdot X(x)}n^{\alpha}\nabla_{\alpha}f(x)\; -\; f(x)n^{\alpha}\nabla_{\alpha}e^{-ik\cdot X(x)}\; ]
\end{equation}
where f(x) = f(X(x)), $n^{\alpha}$ is a unit normal to the embedding $(X^{+}(x),X^{-}(x))$, and in fact $\sqrt{g}n^{\alpha}\nabla_{\alpha}f(x)\ =\ \pi_{f}(x)$. Thus given (f(x),$\pi_{f}(x)$) on a spatial slice, we can obtain $(a_{k}, a^{*}_{k})$.\\
\begin{equation}
\begin{array}{lll}
a_{k}\ =\ \int \sqrt{g}\ [\ u_{k}^{*}n^{\alpha}\nabla_{\alpha}f\ -\ fn^{\alpha}\nabla_{\alpha}u_{k}^{*}\ ]\\
\vspace*{0.1in}
\quad =\ \int[\; u_{k}^{*}\pi_{f}\; -\; \sqrt{g}f\; (\; -ik^{-}\sqrt{\frac{X^{+'}}{X^{-'}}}\; +\; ik^{+}\sqrt{\frac{X^{-'}}{X^{+'}}}\; )]\\
\vspace{0.1in}
\quad =\ \int [\pi_{f}\ +\ ik^{-}X^{+'}f\ -\ ik^{+}X^{-'}f]
\end{array}
\end{equation}
Where we have used $n^{+}\ =\ \sqrt{\frac{X^{+'}}{X^{-'}}}$ , $n^{-}\ =\ -\sqrt{\frac{X^{-'}}{X^{+'}}}$ and $\sqrt{g}\ =\ \sqrt{X^{+'}X^{-'}}$.\\
Using $k^{\pm}\ =\ \frac{1}{2}(k\ \pm \vert k\vert)$ we can show that,\\
\begin{equation}
\begin{array}{lll}
a_{k}\ =\ \int e^{-ikX^{-}}Y^{-}\quad\quad	k>0\\
\vspace*{0.1in}
a_{k}\ =\ \int e^{-ikX^{+}}Y^{+}\quad\quad	k<0\\
\vspace*{0.1in}
a_{0}\ =\ \int \pi_{f}
\end{array}
\end{equation}
$a_{k}^{*}$ are defined by complex conjugating the $a_{k}$.\\
By explicit calculations one can check that these functions are Dirac observables.\\
Their Poisson algebra is given by,\\
\begin{equation}
\begin{array}{lll}
\lbrace a_{k},a_{l}\rbrace \ =\ 0.\\
\vspace*{0.1in}
\lbrace a_{k}^{*},a_{l}^{*}\rbrace\ =\ 0.\\
\vspace*{0.1in}
\lbrace a_{k},a_{l}^{*}\rbrace\ =\ \vert k\vert\delta(k,l)
\end{array}
\end{equation}

\subsection{Quantization of $a_{k}$}

\hspace*{0.2in} In this section we show how to promote $(a_{k},a_{k}^{*})$ to densely defined operators on $\mathcal{H}_{phy}$. This prescription can at best be viewed as an ad-hoc way of trying to promote regulated expressions from $\mathcal{H}_{kin}$ to $\mathcal{H}_{phy}$. We hope that a better scheme for doing this emerges in future or that the one given here is more justified.\\
\hspace*{0.2in} Given a (strong) Dirac observable (one that strongly commutes with the Virasoro constraints), ideal way to promote it to an operator on $\mathcal{H}_{phy}$ is as follows. One first defines an operator on $\mathcal{H}_{kin}$ and if this operator is G-equivariant(where G here is the direct product of 2 copies of diffeomorphisms acting on $\mathbf{R}$), then one can define an operator on $\mathcal{H}_{phy}$ simply by dual action. We will show how this procedure fails here \cite{th3}. (This is analogous to a generic problem in LQG of defining connection dependent operators on $\mathcal{H}_{diff}$.)\\
for $ k\ >\ 0$,\\
\begin{equation}
a_{k}\ =\ \int e^{ikX^{-}}Y^{-}
\end{equation}
\hspace*{0.2in} In order to represent $a_{k}$ on $\mathcal{H}_{kin}$ we have to triangulate our spatial slice $\Sigma$ by 1-simplices (closed intervals). Let T be a triangulation of $\sigma$. Given a state $f^{-}_{s}$ for the left moving sector, we choose a triangulation $T(\gamma(s))$ adapted to $\gamma(s)$ i.e. the triangulation is such that all the vertices of $\gamma(s)$ are vertices of $T(\gamma(s))$ . Classically we know that,\\
\begin{equation}
\frac{h_{\triangle_{m}}(Y^{-})\, -\, h_{\triangle_{m}^{-1}}(Y^{-})}{\vert \triangle_{m}\vert}\, =\, Y^{-}(\frac{v_{m}\, +\, v_{m+1}}{2})\, +\, O((\vert \triangle_{m}\vert)^{2}).
\end{equation}
Where $\triangle_{m}\in T(\gamma(s))$ (It is a closed interval in say Cartesian co-ordinate system), and ($v_{m}$,$v_{m+1}$) are beginning and terminating vertices of $\triangle_{m}$ respectively.\\
Now we can write $a_{k}$ as the limit of a Riemann sum,\\
\begin{equation}
a_{k}\, =\, lim_{T \rightarrow \Sigma}\, a_{k,T(\gamma(s))}
\end{equation}
where
\begin{equation}\label{eq:confused}
a_{k,T(\gamma(s))}\, =\, \sum_{\triangle_{m}\in T(\gamma(s))} e^{ik\hat{X}^{-}(v_{m})}[h_{\triangle_{m}}Y^{-}\ -\ h_{\triangle_{m}^{-1}}(Y^{-})].
\end{equation}
$a_{k,T(\gamma(s))}$ can be represented on $H_{kin}$ as follows.
\begin{displaymath}
\hat{a}_{k,T(\gamma(s))}f^{-}_{s}\ =\ \sum_{\triangle_{m}\in T(\gamma(s))} e^{ik\hat{X}^{-}(v_{m})}[h_{\triangle_{m}}\ -\ h_{\triangle_{m}^{-1}}]\, f^{-}_{s}
\end{displaymath}
\hspace*{0.2in} Similar expression holds for $k\, <\, 0$ with ($X^{-}$, $Y^{-}$) replaced by ($X^{+}$, $Y^{+}$) and the resulting operator acting on $f^{+}_{s}$. Also one can define $\hat{a}_{k,T(\gamma(s))}^{\dagger}$ using the inner product on $\mathcal{H}_{kin}$.\\
\hspace*{0.2in} At finite triangulation (i.e. when number of simplices in T are finite) $\hat{a}_{k,T(\gamma(s))}$ is \emph{not} Virasoro-equivariant,\\
\begin{equation}
\begin{array}{lll}
\hat{U}(\phi^{-})a_{k,T(\gamma(s))}\hat{U}^{-1}(\phi^{-})\, =\, \hat{U}(\phi^{-})\sum_{\triangle_{m}\in T(\gamma(s))}e^{ik\hat{X}^{-}(v_{m})}[h_{\triangle_{m}}\, -\, h_{\triangle_{m}^{-1}}]\, \hat{U}^{-1}(\phi^{-})\\
\vspace*{0.2in}
=\sum_{\triangle_{m}}\hat{U}(\phi^{-})e^{ik\hat{X}^{-}(v_{m})}\hat{U}^{-1}(\phi^{-})\, \hat{U}(\phi^{-})[h_{\triangle_{m}}\, -\, h_{\triangle_{m}^{-1}}]\hat{U}^{-1}(\phi^{-})\\
\vspace*{0.1in}
=\sum_{\triangle_{m}}e^{ikX^{-}(\phi^{-1}(v_{m}))}\, [h_{\phi^{-1}(\triangle_{m})}\, -\, h_{\phi^{-1}(\triangle_{m}^{-1})}]\\
\vspace*{0.1in}
=\sum_{\overline{\triangle_{m}}\in \phi(T(\gamma(s)))}e^{ik\hat{X}^{-}(\overline{v}_{m})}[\, h_{\overline{\triangle}_{m}}\, -\, h_{\overline{\triangle}_{m}^{-1}}\, ]\\
\vspace*{0.1in}
=\, a_{k,\phi(T(\gamma(s)))}.
\end{array}
\end{equation}
\hspace*{0.2in} Thus, $\hat{a}_{k,T(\gamma(s))}$ cannot be promoted to an operator on $\mathcal{H}_{phy}$ simply by dual action. This problem was also encountered by Thiemann in \cite{th3}. As argued by him, if we try to remove the triangulation by taking the continuum limit then we either get zero(in weak operator topology) or infinity(in strong operator topology).\\
There are two ways to get around this problem. First way is due to Thiemann.\\
\hspace*{0.2in} There the idea was to use the graph underlying a state itself as a triangulation, and define a strongly Virasoro-invariant operator on $\mathcal{H}_{kin}$. Here we propose a different way.\footnote{The idea of defining regulated operators on $\mathcal{H}_{phy}$ in this way was suggested to us by Madhavan Varadarajan.} Essentially we use a sort of gauge-fixing in the space of (diff) equivalence class of charge-networks (defined as the triple $(\gamma ,\; \vec{l}(\gamma),\; \vec{k}(\gamma))$) to define an operator corresponding to $a_{k}$ on $\mathcal{H}_{phy}$. As will be argued later, Thiemann's proposal can be considered as a special case of ours.\\
\hspace*{0.2in} Given an orbit of diffeomorphism equivalence class of charge-networks, $[s]\; =\; \lbrace\phi\cdot s\vert\; \phi\in Diff\Sigma \rbrace$
we fix \underline{once and for all} a network $s_{0}\; =\; (\gamma_{0},\; \vec{l}(\gamma),\; \vec{k}(\gamma))$ and a triangulation $T(\gamma_{0}(s_{0}))$ adapted to it. Now for any s in the orbit, such that $s\; =\; \phi\cdot s_{0}\; =\; (\phi^{-1}(\gamma_{0}),\; \vec{l}(\gamma),\; \vec{k}(\gamma))$ we choose the corresponding triangulation $T(\gamma(s))$ such that $\hat{a}_{T(\gamma(s))}\; =\; \hat{U}(\phi)\hat{a}_{T(\gamma_{0}(s_{0}))}\hat{U}^{-1}(\phi)$. Now let $\Psi \in \mathcal{H}_{phy}$. One can show that this family of operators are cylindrically consistent and define a operator on $\mathcal{H}_{kin}$. The resulting operator on $\mathcal{H}_{phy}$ defined by the dual action turns out to be densely defined.\\
\begin{equation}
(\hat{a}_{k'}\Psi)[f_{s}]\, =\, \Psi[\, \hat{a}_{k,T(\gamma_{0}
)}^{\dagger}\, f_{s_{0}}\, ]
\end{equation}
\hspace*{0.2in} Where for the sake of simplicity we have suppressed $\pm$ labels indicating left (right) moving sectors. Here as defined earlier $\gamma_{0}(s_{0})$ is the graph which is fixed in the orbit of $\gamma(s)$, and $T(\gamma_{0}(s_{0}))$ is a fixed triangulation adapted to it. This proposal ,of defining $\hat{a_{k'}}$ on $\mathcal{H}_{phy}$, is as we emphasized earlier rather ad-hoc as it involves an arbitrary choice $s_{0}$ and triangulation $T(\gamma_{0}(s_{0}))$. It nonetheless results in a ``regulated'' and densely defined operator on $\mathcal{H}_{phy}$.\\
\hspace*{0.2in} We will now argue that Thiemann's proposal of defining a Virasoro-invariant operator directly on $\mathcal{H}_{kin}$ ( \cite{th3} pg.28 ) can be subsumed by the prescription given above. Note that in \cite{th3} spatial topology is compact ($S^{1}$), whence we have to modify the proposal given there accordingly as in our case the spatial manifold is $\mathbb{R}$. Let us first note how Thiemann's prescription applies to our perennials.\\
1. Choose the graph underlying a state itself as a triangulation, by adding fiducial edges if necessary.\\
2.Then the operator (in our case $\hat{a}_{k,\gamma(s)}$) acting on a basis-state $f_{s}$ results in a linear combination of states $f_{s'}$ such that $\gamma(s')\subset \gamma(s)$. Thus,
\begin{equation}
\hat{U}(\phi)\; \hat{a}_{k,\gamma(s)}f_{s}\; =\; \sum\; b_{I}\hat{U}(\phi)f_{s_{I}}\; =\; \sum\; b_{I}f_{\phi(s_{I})}.
\end{equation}
Using (\ref{eq:confused}) one can write this more explicitly as,\\
\begin{equation}\label{eq:jane}
\begin{array}{lll}
\; \hat{U}(\phi)\; (\sum_{e\in E(\gamma)}\; e^{ik\hat{X}^{-}(b(e))}\; [h_{e}(Y^{-})\ -\ h_{e^{-1}}(Y^{-})])\; f_{s}\\
\vspace*{0.1in}
=\; \hat{U}(\phi)\; (\sum_{e\in E(\gamma)}\; e^{ik\frac{1}{2}(k_{e}^{-}+k_{e-1}^{-})}\; e^{i\hbar\alpha(e,\gamma)}[f_{s'}\; -\; f_{s''}])\\
\vspace*{0.1in}
=\; \sum_{e\in E(\gamma)}\; e^{ik\frac{1}{2}(k_{e}^{-}+k_{e-1}^{-})}\; e^{i\hbar\alpha(e,\gamma)}[f_{\phi\cdot s'}\; -\; f_{\phi\cdot s''}]\\
\vspace*{0.1in}
=\; \sum_{e\in E(\phi^{-1}(\gamma))}\; e^{ik\frac{1}{2}(k_{e}^{-}+k_{e-1}^{-})}\; e^{i\hbar\alpha(e,\phi^{-1}(\gamma))}[f_{\phi\cdot s'}\; -\; f_{\phi\cdot s''}]
\end{array}
\end{equation}
where $s'\; =\; (\gamma,\; ((k_{1},l_{1}),...,(k_{e},l_{e}+1),...,(k_{N},l_{N})))$ and $s''\; =\; (\gamma,\; ((k_{1},l_{1}),...,(k_{e},l_{e}-1),...,(k_{N},l_{N})))$. and in the last line we have made use of the fact that
$(k_{\phi^{-1}(e)}, l_{\phi^{-1}(e)})\; =\; (k_{(e)}, l_{(e)})$.\\
However it is easy to convince oneself that the last line in (\ref{eq:jane}) equals,\\
\begin{equation}
\hat{a}_{k,\gamma(\phi\cdot s)}\; f_{\phi\cdot s}\; =\; \hat{a}_{k,T}\; \hat{U}(\phi)\; f_{s}.
\end{equation}
This shows Virasoro equivariance of $\hat{a}_{k,\gamma(s)}$.\\
\hspace*{0.2in} The above proof crucially relies on the fact that the triangulation used to regulate the operator is same as the graph (underlying the state on which the operator acts) itself. We now show how to achieve this by adding fiducial edges to the graph This is where Thiemann's prescription has to be slightly modified as in \cite{th3} spatial topology is that of $S^{1}$.\\
\hspace*{0.2in} Given any basis-state $f_{s}$ we can always write it as a state $\tilde{f}_{\tilde{s}}\in\; \mathcal{H}_{kin}$ such that
$\tilde{\gamma}(\tilde{s})\; =\; e_{L}\cup\; \gamma(s)\cup\; e_{R}$, where $e_{L}$ and $e_{R}$ are edges from $-\infty$ to initial vertex of $\gamma(s)$ and from final vertex of $\gamma(s)$ to  $\infty$ respectively (See figure below). In fact we can define a new basis for $\mathcal{H}_{kin}$ as follows. Any element of the basis $f_{\tilde{s}}$ is defined to be based on a graph which is of the form $\tilde{\gamma}(\tilde{s})\; =\; e_{L}\cup\; \gamma\cup\; e_{R}$ where $\gamma$ is a subgraph of $\tilde{\gamma}(\tilde{s})$ such that $e_{L}$ and $e_{R}$ are as defined above. The charge-pairs $(k_{e_{L/R}},\; l_{e_{L/R}})$ are allowed to be (0,0) but $(k_{e_{1}},l_{e_{1}})$ and $(k_{e_{N}},l_{e_{N}})$ are not allowed to be (0,0).
( Here $e_{1}$ and $e_{N}$ are initial and final edges of $\gamma$ respectively.)\footnote{ The introduction of new basis is only to show how Thiemann's prescription is consistent with ours and will not be used in the rest of the paper anywhere.}
Thus the graphs on which the new basis is defined itself becomes triangulation of $\Sigma$ and Thiemann's prescription follows.\\
\hspace*{0.2in} How does our definition of $\hat{a'}_{k}$ subsume Thiemann's definition as a special case? The answer is as follows. Once we choose an $s_{0}$ in the orbit of s, choose $T(\gamma_{0}(s_{0}))\; =\; e_{L}\cup \gamma_{0}\cup e_{R}$ as shown in the figure.\\
\begin{center}
\includegraphics[scale=0.7]{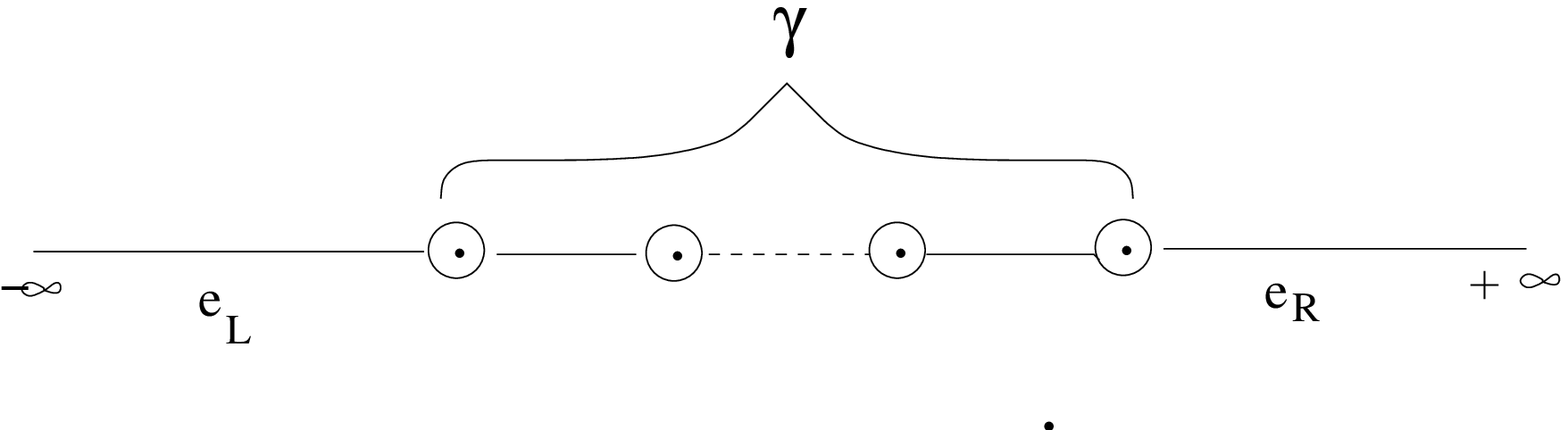}
fig.1
\end{center}
\hspace*{0.2in} The resulting operator $\hat{a}^{\dagger}_{T(\gamma_{0}(s_{0}))}$ is Virasoro invariant on $\mathcal{H}_{kin}$.
\hspace*{0.2in} Whence $\hat{a'}_{k}$ is the dual of a linear operator $\hat{a}^{\dagger}_{T}$  obtained on $\mathcal{H}_{kin}$ via cylindrical consistency.\\
\hspace*{0.2in} We now show that the adjoint of $a_{k'}$ defined on $\mathcal{H}_{phy}$ using the inner product is consistent with the definition,\\
\begin{equation}
(\hat{a}_{k'}^{*}\Psi)[f_{s}]\, =\, \Psi[\, \hat{a}_{k,T(\gamma_{0}(s_{0}))}\, f_{s_{0}}\, ].
\end{equation}
Let $\Psi_{1}\; ,\; \Psi_{2}\;  \in \mathcal{H}_{phy}$, then\\
\begin{equation}
\begin{array}{lll}
<\; \Psi_{1},\; \hat{a'}^{*}\Psi_{2}\; >\; & = & <\; \Psi_{1}\hat{a'},\; \Psi_{2}\; >\\
\vspace*{0.1in}
& = & <\; \Psi_{2},\; \hat{a'}\Psi_{1}\; >^{*}\\
\vspace*{0.1in}
& = & [\; \hat{a'}\Psi_{1}(f_{s_{2}})\; ]^{*}\\
\vspace*{0.1in}
& = & [\; \Psi_{1}(\hat{a}^{\dagger}_{T(\gamma_{2})}f_{s_{2}})\; ]^{*}\\
\vspace*{0.1in}
& = & [\; \sum_{\phi}<f_{s_{1}}\hat{U}(\phi)\; ,\; \hat{a}^{\dagger}_{T(\gamma_{2})}f_{s_{2}}>\; ]^{*}\\
\vspace*{0.1in}
& = &\sum_{\phi}<\; f_{s_{2}},\; \hat{U}(\phi)\; \hat{a}_{T(\gamma_{1})}\;\hat{U}(\phi)^{-1}f_{\phi(s_{1})}\; >\\
\vspace*{0.1in}
& = &\sum_{\phi}<\; f_{s_{2}},\; \hat{U}(\phi)\; \hat{a}_{T(\gamma_{1})}f_{(s_{1})}\; >\\
\vspace*{0.1in}
& = &\sum_{\phi^{-1}}<\; f_{s_{2}}\; \hat{U}(\phi^{-1})\; ,\; \hat{a}_{T(\gamma_{1})}f_{s_{1}}\; >\\
\vspace*{0.1in}
& = &\Psi_{2}(\; \hat{a}_{T(\gamma_{1})}f_{s_{1}}\; )
\end{array}
\end{equation}
where $\eta(f_{s_{I}})\; =\; \Psi_{I}$ for I = 1,2 and we have suppressed the $\pm$ indices for clarity.

\subsection{Commutation relations}

\hspace*{0.2in} Next we study the commutator algebra generated by the Dirac observables $(a_{k},a_{k}^{*})$ in quantum theory. Contrary to the classical Poisson algebra which closes, we show that in the quantum theory even $(a_{k}\, ,\, a_{l})$ do not in general commute with each other. It is plausible that this will have serious implications on causal structure of the quantum theory and the issue is far from being resolved. Recall that the physical content of parametrized free field theory (at least classically) is same as that of ordinary free field theory on flat spacetime. Whence we could have started with the reduced phase space co-ordinatized by $(a_{k},a_{k}^{*})$, and its representation on Fock space will result in a quantum theory in which fields separated by spacelike interval will commute. Also the two point functions will decay exponentially outside the light cone. However \emph{if} the commutator algebra of $(a_{k},a_{k}^{*})$ gets deformed in the quantum theory then it is \emph{not} clear in what sense the causal structure defined by the background spacetime is preserved. In fact as we are not aware of a state (or a class of states) in $\mathcal{H}_{phy}$ which correspond to the Fock vacuum, it is not even known how to define two point functions. (using which we can study causal relations.)\\
%It is not clear what are the implications of this result on the causal structure of the quantum theory. (This question is presumably not too trivial to answer as we are not aware of a state(or class of states) in $\mathcal{H}_{phy}$ which correspond to the Fock vacuum( anhilated by $a_{k}$ )using which one can define 2 pt. functions and study causal relations.)\\%
\\
\underline{The commutator}\\
\hspace*{0.2in} When $k < 0$ and $l > 0$ it is clear that $[\hat{a}_{k'}\, ,\, \hat{a}_{l'}]$ will be trivially zero as $\hat{a}_{k'}$ acts on right-moving sector ($\mathcal{H}_{phy}^{+}$) and $\hat{a}_{l'}$ acts on left-moving sector ($\mathcal{H}_{phy}^{-}$) whence they commute.\\
Let us consider the case when k,l $<$ 0. Remaining case ( k,l $>$ 0) can be handled similarly.\\
As $a_{k}\, =\, \int Y^{+}(x)e^{ik\, X^{+}(x)}$, we only look at the right-moving(+) sector of $\mathcal{H}_{phy}$.\\
\begin{equation}
\begin{array}{lll}
(\, [\hat{a}_{k'}\, ,\, \hat{a}_{l'}]\Psi^{+}\, )f^{+}_{s}\, =\, ((\hat{a}_{k'}\hat{a}_{l'}\, -\, \hat{a}_{k'}\hat{a}_{l'})\Psi^{+}\, )f^{+}_{s}\\
\vspace*{0.1in}
=(\hat{a}_{k'}\hat{a}_{l'}\Psi^{+}\, )f^{+}_{s}\, -\, (\hat{a}_{l'}\hat{a}_{k'}\Psi^{+}\, )f^{+}_{s}\\
\vspace*{0.1in}
=(\hat{a}_{l'}\Psi^{+}\, )(\hat{a}_{k,T(\gamma_{0}(s_{0}))}^{\dagger}f^{+}_{\gamma_{0}})\, -\, (\hat{a}_{k'}\Psi^{+}\, )(\hat{a}_{l,T(\gamma_{0}(s_{0}))}^{\dagger}f^{+}_{s_{0}})
\end{array}
\end{equation}
\hspace*{0.2in} Where $f^{+}_{s}$ is an arbitrary state in the kinametical Hilbert space of the right moving sector $\mathcal{H}_{kin}^{+}$ and
as before $s_{0}$ is a fixed charge-network in the orbit of $s$. Let us look at both the terms separately.\\
\underline{Term 1 - $(\, \hat{a}_{l'}\Psi^{+}\, )(\, \hat{a}_{k,T(\gamma_{0}(s_{0}))}^{\dagger}\, f^{+}_{s_{0}}\, )$}\\
\hspace*{0.2in} Now we employ a specific choice of triangulation $T(\gamma_{0}(s_{0}))$. This choice is motivated by the requirement of simplicity. We will argue shortly that the result(at least qualitatively) does not depend on this particular choice.\\
So let us choose $T(\gamma_{0}(s_{0}))$ = $\gamma_{0}\, \cup e_{L}\, \cup e_{R}$. where $e_{L}$ and $e_{R}$ are as shown in the figure 1.\\
(Remark : With this choice of the triangulation the continuum limit is approached only when $\vert E(\gamma_{0}) \vert$ tends to $\infty$.)\\
so,\\
\begin{equation}
\hat{a}_{k,T(\gamma_{0}(s_{0}))}^{\dagger}\, =\, \frac{1}{2i}\sum_{e_{I}\in (\gamma_{0}\cup e_{L}\cup e_{R})}e^{-ik\hat{X}^{+}(v_{I})}\, [\, h_{e_{I}}\, -\, h_{e_{I}^{-1}}\, ]
\end{equation}
Here $v_{I}$ = $b(e_{I})$.\\
Similarly we choose $T(\gamma_{0}(s_{0}))$ = $\gamma_{0} \cup e_{L} \cup e_{R}$, which implies,\\
\begin{equation}
\Psi(\, \hat{a}_{l,T(\gamma_{0}(s_{0}))}^{\dagger}\, \hat{a}_{k,T(\gamma_{0}(s_{0}))}^{\dagger}\, f_{\gamma_{0}}^{\dagger}\, )\, =\,
-\frac{1}{4}\, \Psi\, (\, \sum_{e_{J}}\, e^{-il\hat{X}^{+}(v_{J})}[\, h_{e_{J}}\, -\, h_{e_{J}^{-1}}\, ]\, \sum_{e_{I}}e^{-ik\hat{X}^{+}(v_{I})}[\, h_{e_{I}}\, -\, h_{e_{I}^{-1}}\, ]\, )\\
\end{equation}
Second term $(\, \hat{a}_{k'}\Psi^{+}\, )(\, \hat{a}_{l,T(\gamma_{0}(s_{0}))}^{\dagger}\, f^{+}_{s_{0}}\, )$ can be evaluated similarly and we get,\\
\begin{equation}
(\, [\hat{a}_{k'}\, ,\, \hat{a}_{l'}]\Psi\, )(f^{+}_{s})\, =\, \Psi\, (\, \sum_{e_{J},e_{I}}e^{-il\hat{X}^{+}(v_{J})}e^{-ik\hat{X}^{+}(V_{I})}[( h_{e_{J}}\, -\, h_{e_{J}^{-1}}\, )\, ,\, (\, h_{e_{I}}\, -\, h_{e_{I}^{-1}}\, )]\, )\\
\end{equation}
\hspace*{0.2in} In the above double sum only those edges $(e_{I}\, \, e_{J})$ contribute for which $e_{I}\cap e_{J}\, \neq 0$.\\
Consider the following 2 pairs (I= M,J=M+1) and (I+M+1,J=M) for some fixed M.\\
The contribution to the commutator coming from the above pairs is,\\
\begin{equation}\label{eq:sergio}
\begin{array}{lll}
-\frac{1}{4}\Psi &(& e^{-il\hat{X}^{+}(v_{M+1})}e^{-ik\hat{X}^{+}(v_{M})}[\, (h_{e_{M+1}}\, -\, h_{e_{M+1}^{-1}})\, \, (h_{e_{M}}\, -\, h_{e_{M}^{-1}})\, ]\\
\vspace*{0.1in}
&+&\, e^{-ik\hat{X}^{+}(v_{M+1})}e^{-il\hat{X}^{+}(v_{M})}[\, (h_{e_{M}}\, -\, h_{e_{M}^{-1}})\, \, (h_{e_{M+1}}\, -\, h_{e_{M+1}^{-1}})\, ]\, f^{+}_{s_{0}}\, )\\
\vspace*{0.1in}
&=& -\frac{1}{4}\Psi\, (\, [\, (h_{e_{M+1}}\, -\, h_{e_{M+1}^{-1}})\, ,\, (h_{e_{M}}\, -\, h_{e_{M}^{-1}})\, ]\\
\vspace*{0.1in}
& &\hspace*{0.5in} (\, e^{-il\hat{X}^{+}(v_{M+1})}e^{-ik\hat{X}^{+}(v_{M})}\, -\, e^{-ik\hat{X}^{+}(v_{M+1})}e^{-il\hat{X}^{+}(v_{M})}\; )f^{+}_{s_{0}})
\end{array}
\end{equation}
Now using (\ref{eq:2}) one can show that,\\
\begin{equation}
[h_{e_{I}},\; h_{e_{J}}]\; =\; sin(-i\hbar \alpha(e_{I},e_{J}))\; h_{e_{I}+e_{J}}
\end{equation}
\hspace*{0.2in} It is now straight-forward to evaluate the commutators in the (\ref{eq:sergio}). Whence given a pair of successive edges which lie within the graph $(e_{I}\, ,\, e_{I+1})$ (I = 1,...N-1) their contribution to $(\, [\hat{a}_{k'}\, ,\, \hat{a}_{l'}]\Psi\, )(f^{+}_{s})$ is,\\
\begin{equation}
\begin{array}{lll}
-\frac{1}{4}\Psi\; [\; sin(\frac{1}{2}\hbar i) \sum_{I=1}^{N-1}\, (h_{e_{I}+e_{I+1}}\, +\, h_{e_{I^{-1}}+e_{I+1^{-1}}}\, +\, h_{e_{I^{-1}}+e_{I+1}}\, +\, h_{e_{I}+e_{I+1^{-1}}})\\
\vspace*{0.1in}
\hspace*{1.3in} [\, e^{-il\hat{X}^{+}(v_{I+1})}e^{-ik\hat{X}^{+}(v_{I})}\, -\, e^{-ik\hat{X}^{+}(v_{I+1})}e^{-il\hat{X}^{+}(v_{I})}\, ]f^{+}_{s_{0}}\\
\vspace*{0.1in}
=-\frac{1}{4}\Psi\, [\, sin(\frac{1}{2}\hbar i)\sum_{I=1}^{N-1} ( h_{e_{I}+e_{I+1}}\, +\, h_{e_{I^{-1}}+e_{I+1^{-1}}}\, +\, h_{e_{I^{-1}}+e_{I+1}}\, +\, h_{e_{I}+e_{I+1^{-1}}})\\
\vspace*{0.1in}
\hspace*{1.0in} [\; e^{-\frac{1}{2}i\hbar l(k_{e_{I}}+k_{e_{I+1}})}e^{-\frac{1}{2}i\hbar k(k_{e_{I-1}}+k_{e_{I}})}\, -\, e^{-\frac{1}{2}i\hbar k(k_{e_{I}}+k_{e_{I+1}})}e^{-\frac{1}{2}i\hbar l(k_{e_{I-1}}+k_{e_{I}})}\; ]f^{+}_{s_{0}}
\end{array}
\end{equation}
\hspace*{0.2in} Where we have used $\hat{X}^{+}(v_{I})f^{+}_{s_{0}}\ =\ \frac{1}{2}\hbar (k^{+}_{e_{I}}\ + \ k^{+}_{e_{I+1}})f^{+}_{\gamma_{0}}$ and defined $k_{e_{0}}\ =\ 0$.\\
Finally there are contributions from the pair $(e_{L},e_{1})$ and $(e_{N},e_{R})$,\\
\begin{displaymath}
\begin{array}{lll}
-\frac{1}{4}\Psi\, (\, (e^{-ilk_{e_{1}}}\, -\, e^{-ikk_{e_{1}}})\; sin(\frac{1}{2}\hbar i)\; [h_{e_{1}+e_{L}}\, +\, h_{e_{1}^{-1}+e_{L}^{-1}}\, +\, h_{e_{1}^{-1}+e_{L}}\, +\, h_{e_{1}+e_{L}^{-1}}]f^{+}_{s_{0}}\, )\\
\vspace*{0.1in}
-\frac{1}{4}\Psi\, (\, (e^{-il\hbar k_{e_{N}}}e^{-i\frac{1}{2}k\hbar (k_{e_{N-1}}+k_{e_{N}})}\, -\, e^{-ik\hbar k_{e_{N}}}e^{-i\frac{1}{2}l\hbar (k_{e_{N-1}}+k_{e_{N}})})\\
\vspace*{0.1in}
\hspace*{1.0in} sin(\frac{1}{2}\hbar i)\; [h_{e_{R}+e_{N}}\, +\, h_{e_{R^{-1}}+e_{N^{-1}}}\, +\, h_{e_{R^{-1}}+e_{N}}\, +\, h_{e_{R}+e_{N^{-1}}}]f^{+}_{s_{0}}\, )
\end{array}
\end{displaymath}
Let $e_{L}=e_{0}$ and $e_{R}=e_{N+1}$ , $k_{e_{0}}=k_{e_{N+1}}=0$ we finally get,\\
\begin{equation}
\begin{array}{lll}
([\hat{a}_{k'}\, \, \hat{a}_{l'}]\Psi)(f^{+}_{s})\, & = &\\
\vspace*{0.1in}
&  -\frac{1}{4}\Psi\biggl(sin(\frac{1}{2}\hbar i)\; \sum_{I=0}^{N} [h_{e_{I}+e_{I+1}}\, +\; h_{e_{I^{-1}}+e_{I+1^{-1}}}\, +\; h_{e_{I^{-1}}+e_{I+1}}\, +\; h_{e_{I}+e_{I+1^{-1}}}]&  \\
\vspace*{0.1in}
& (\; e^{-i\frac{1}{2}\hbar l (k_{e_{I}}+k_{e_{I+1}})}e^{-i\frac{1}{2}\hbar k(k_{e_{I-1}}+k_{e_{I}})}\, -\, e^{-i\frac{1}{2}\hbar k(k_{e_{I}}+k_{e_{I+1}})}e^{-i\frac{1}{2}\hbar l(k_{e_{I-1}}+k_{e_{I}})}\, )f^{+}_{s_{0}}\biggr)&
\end{array}
\end{equation}
\hspace*{0.2in} Thus it is clear that in general the commutator $[\hat{a}_{k'}\, \, \hat{a}_{l'}]$ (k , l $<$ 0)does not vanish on $\mathcal{H}_{phy}$. The commutator for $[\hat{a}_{k'}\, \, \hat{a}_{l'}]$ with (k , l $>$ 0) is exactly similar with all operators acting on the left moving sector.\\
\hspace*{0.2in} Now we give a heuristic proposal showing existence of (a class of) states  on which the commutator $[\hat{a}_{k'}\, \, \hat{a}_{l'}]$ vanishes. Ideally one would like to do a semi-classical analysis of the expectation value of the commutators to see if the non-zero contributions are sub-leading. This is an open question that we have not addressed in the present paper. In what follows we argue for the existence of states, possibly in ITP(infinite tensor product extension \cite{th4}) of $\mathcal{H}_{phy}$ on which the commutator vanishes.\\
\hspace*{0.2in} Notice that given a $\Psi$ $\in$ $\mathcal{H}_{phy}$,  $([\hat{a}_{k'}\, \, \hat{a}_{l'}]\Psi)(f^{+}_{s})$ is non-zero iff the ``embedding-component'' of $\Psi$ is group averaged distribution obtained from the ``embedding-component'' of $f^{+}_{s}$. In other words if $\Psi\ =\ \vert 2N+1,2N+2,([-N,l_{1}] ,...,[N,l_{2N}])\ >$ where the matter-charges $(l_{1},...l_{2N})$ are arbitrary but non-zero then  $([\hat{a}_{k'}\, \, \hat{a}_{l'}]\Psi)(f^{+}_{s})$  is non-zero iff $\vert E(\gamma)\vert$ = 2N+1 , the embedding charges on the edges of $\gamma$ constitute the set (-N,...,N) and the matter charges form a set $(l_{1},...,l_{I}\pm 1,l_{I+1}\pm 1,...l_{2N})$  for some I.\\
\begin{equation}
\begin{array}{lll}
([\hat{a}_{k'}\, \, \hat{a}_{l'}]\Psi)(f^{+}_{s})\, & = &\\
\vspace*{0.1in}
& & -\frac{1}{4}\Psi\, sin(\frac{1}{2}\hbar i)\; \biggl(\, \sum_{I=0}^{N} [\; h_{e_{I}+e_{I+1}}\, +\, h_{e_{I^{-1}}+e_{I+1^{-1}}}\, +\, h_{e_{I^{-1}}+e_{I+1}}\, +\, h_{e_{I}+e_{I+1^{-1}}}\; ]  \\
\vspace*{0.1in}
& & \sum_{n=-N}^{N}\; {\bf [}e^{-i\hbar (l+k)n}e^{-\frac{1}{2}i\hbar (l-k)}\; -\; e^{-i\hbar (l+k)n}e^{-\frac{1}{2}i\hbar (k-l)}{\bf ]}f^{+}_{s_{0}}\ \biggr)
\end{array}
\end{equation}
\hspace*{0.2in} Now as $N \rightarrow \infty$ and each $e_{I}$ shrinks to it is vertex $v_{I}$, and if we assume that to leading order in $\frac{1}{N}$,  $h_{e_{I}} \rightarrow\; 1$ then one gets,\\
\begin{equation}
\begin{array}{lll}
([\hat{a}_{k'}\, \, \hat{a}_{l'}]\Psi)(f^{+}_{s})\, & = &\\
\vspace*{0.1in}
& & -\frac{1}{4}\Psi\, \biggl(\; \sum_{n \in \bold{Z}}\; {\bf [}e^{-i\hbar (l+k)n}e^{-\frac{1}{2}i\hbar (l-k)}\; -\; e^{-i\hbar (l+k)n}e^{-\frac{1}{2}i\hbar (k-l)}{\bf ]}f^{+}_{s_{0}}\ \biggr)\\
\vspace*{0.1in}
& = & -\frac{1}{4}\Psi\, \biggl(\; \delta(l+k) sin(\frac{1}{2}\hbar (l-k))\; f^{+}_{s_{0}}\biggr)
\end{array}
\end{equation}
which equals 0 for l,k $<$ 0.\\
Couple of comments are in order :\\
1. We have not displayed semi-classicality in the sense that we have not shown that the
non-vanishing terms in $[\hat{a}_{k'}\, \, \hat{a}_{l'}]$ are sub-leading corrections on a class of states in $\mathcal{H}_{phy}$.\\
2. The above result does not depend on our choice of triangulation $T(\gamma_{0}(s_{0}))\; =\; \gamma_{0}\cup e_{L}\cup e_{R}$.
Consider any triangulation, T which is adapted to $\gamma_{0}$ in the sense that the vertex set of $\gamma_{0}$ is a subset of the vertex-set of T. Then it can be shown that only those edges which intersect the vertices of the graph contribute. Contributions from all other edges cancel out pairwise.\\
The calculation of $[\hat{a}_{k'}\, \, \hat{a}_{l}^{\dagger'}]$ proceeds similarly.\\
\begin{equation}
\begin{array}{lll}
([\hat{a}_{k'}\, \, \hat{a}_{l}*']\Psi)(f^{+}_{s})\, & = &\\
\vspace*{0.1in}
& & \frac{1}{4}\Psi\; sin(\frac{1}{2}\hbar i)\; \biggl(\, \sum_{I=0}^{N} [\; h_{e_{I}+e_{I+1}}\, +\; h_{e_{I^{-1}}+e_{I+1^{-1}}}\, +\, h_{e_{I^{-1}}+e_{I+1}}\; +\; h_{e_{I}+e_{I+1^{-1}}}]  \\
\vspace*{0.1in}
& ( & e^{i\frac{1}{2}\hbar l(k_{e_{I}}+k_{e_{I+1}})}e^{-i\frac{1}{2}\hbar k(k_{e_{I-1}}+k_{e_{I}})}\, -\, e^{-i\frac{1}{2}\hbar k(k_{e_{I}}+k_{e_{I+1}})}e^{i\frac{1}{2}\hbar l(k_{e_{I-1}}+k_{e_{I}})}\, \, )f^{+}_{s_{0}}\ \biggr)
\end{array}
\end{equation}
\\
which as $N \rightarrow \infty$ and each $e_{I}$ shrinks to it is vertex $v_{I}$ gives,
\\
\begin{equation}
([\hat{a}_{k'}\, \, \hat{a}_{l}^{*'}]\Psi)(f^{+}_{s})\; =  \frac{1}{4\hbar }sin(\frac{\hbar i}{2})\Psi(\; \delta(l-k) sin(\frac{1}{2}\hbar (l+k))\; f^{+}_{\gamma_{0}}\; )
\end{equation}
which is a specific quantum deformation of the classical Poisson bracket.\\
\hspace*{0.2in} Our heuristic calculations show that it is plausible that on a specific class of states with countably infinite edges the commutator algebra generated by ($a^{k'}$, $a^{{k}^{*'}}$ and 1) closes and is a specific deformation of the Poisson algebra. Such states cannot lie in $\mathcal{H}_{phy}$ but in infinite tensor product extension thereof[\cite{th3}].\\
\underline{Note that}, in \cite{th3}  semi-classical states have been defined by using graphs with large but finite number of edges. However, the heuristic calculations displayed above suggest that when spatial slice is non-compact, ideal home for semi-classical states is the ITP extension of $\mathcal{H}_{phy}$.
\section{Quantum dynamics via complete (evolving) observables}
\hspace*{0.2in} In the previous section we obtained a complete set of Dirac observables for our system and showed how this set can be promoted to well defined operators on $\mathcal{H}_{phy}$. As emphasized above, we do not have the representation of the  Poisson algebra of Dirac observables on $\mathcal{H}_{phy}$. The algebra gets deformed and we believe  it is an important open question to know the full quantum algebra. But we can still proceed and  use these perennials to help us tackle the problem of time in our system. Using the concept of complete observables due to Dittrich \cite{bianca}, we will define a notion of dynamics on $\mathcal{H}_{phy}$. We begin by defining complete observables and their dynamics first in classical theory.\\
\hspace*{0.2in} The canonical co-ordinates on the phase-space are $(f(x),\; \pi_{f}(x),\; X^{+}(x),\; \Pi_{+}(x),\; X^{-}(x),\; \Pi_{-}(x))$. In what follows, we will treat $x\in \sigma$ as a label set and think of the canonical fields as functionals on the phase-space labelled by x i.e. $f(x)\; :\; \mathcal{M}\; \rightarrow\; \mathbf{R}$
Now choose the following gauge fixing conditions for the two Virasoro constraints,\\
\begin{equation}
X^{\pm}(x)\; =\; X_{p}^{\pm}(x)\; \forall\; x \in \sigma
\end{equation}
where $X^{\pm}_{p}\; :\; \sigma\; \rightarrow\; M^{2}$ is a prescribed embedding of $\sigma$ in  Minkowski space $M^{2}$.\\
\hspace*{0.2in} It is easy to see that these are good gauge fixing conditions in the sense that they define global gauge slices, i.e. one can draw a gauge orbit passing through any point on the constraint surface which intersects this slice transversally.\\
\hspace*{0.2in} Using the functional on phase-space f(x), for a given x, we can now construct a complete observable $f(x)[X^{+}_{p},X^{-}_{p};m]$ as follows.\\
\hspace*{0.2in} Given a point $m\; =\; (f,\; \pi_{f},\; X^{+},\; X^{-})$ on the constraint surface, and a gauge orbit $\mathcal{G}_{m}$ passing through it, we ask for the value of f(x) at that point on the gauge orbit $m'$ which intersect the gauge slice defined by the above gauge fixing conditions. We define,\\
\begin{equation}
f(x)[X^{+}_{p},X^{-}_{p};m]\; =\; f(x)[m']
\end{equation}
\hspace*{0.2in} It is immediately clear from the definition that $f(x)[X^{+}_{p},X^{-}_{p};m]$ is invariant under gauge transformations, and one can show that \cite{bianca} it has a (weakly) unique extension off the constraint surface.\\
\hspace*{0.2in} A complete observable $\pi_{f}(x)[X^{+}_{p},X^{-}_{p};m]$ corresponding to $\pi_{f}$ can be defined analogously.\\
\hspace*{0.2in} There is an alternate characterization of  complete observables in parametrized field theory which immediately yield the explicit expression for these quantities.\footnote{ The complete observables also satisfy a functional differential equation see \cite{bianca} which in our case can be explicitly solved to get the same expression  for the complete observables that are given here.} Given a maximal classical solution to $\Box f(X)\; =\; 0$ which lies in the gauge orbit $\mathcal{G}$ passing through $(f,\pi_{f},X^{\pm})$ what is the Cauchy data corresponding to it on the slice given by $X^{\pm}_{p}:\; \sigma\; \rightarrow\; M^{2}$. The answer is immediate,\\
\begin{equation}
\begin{array}{lll}
f(x)[X^{+}_{p},X^{-}_{p};m]\; & = & \frac{1}{\sqrt{2\pi}}\int_{-\infty}^{+\infty}\frac{dk}{\vert k\vert}[\; a_{k}(m)e^{ik\cdot X_{p}(x)}\; +\; a_{k}^{*}(m)e^{-ik\cdot X_{p}(x)}]\\ \\
\vspace*{0.2in}
\pi_{f}(x)[X^{+}_{p},X^{-}_{p};m]\;\; & = & \frac{i}{\sqrt{2\pi}}[\; \int_{0}^{\infty}dk\; X_{p}^{-'}(x)a_{k}(m)e^{ik\cdot X_{p}}\; +\; c.c.\\
& & - \int_{-\infty}^{0}dk\; X_{p}^{+'}(x)a_{k}(m)e^{ik\cdot X_{p}}\; -\; c.c.\; ]
\end{array}
\end{equation}
\hspace*{0.2in} Thus given a free scalar field  on flat space-time, its Cauchy data on a prescribed slice gives complete observables of the corresponding parametrized field theory.\\
\hspace*{0.2in} As shown in \cite{bianca}, Poisson bracket of two complete observables is an observable (in fact it is shown to be a complete observable.) Thus the space of all complete observables form an *-Poisson algebra.
\subsection{Classical dynamics}

One can define non-trivial gauge transformations $\hat{\alpha}_{\tau}$
on the space of observables which generalizes Rovelli's idea of evolving constant of motion to arbitrary number of constraints.
The basic idea is to see how does $f(x)[X^{+}_{p},X^{-}_{p};m]$ change when one changes the gauge-fixing slices $X^{\pm}\; =\; X_{p}^{\pm}$ under gauge transformations.\\
In our case these transformations simply amount to changing the parameters $X_{p}^{\pm}(x)$ additively.\\
\begin{equation}\label{eq:one}
X^{\pm}_{p}(x)\; \rightarrow\; X^{\pm}_{p}(x)\; + \tau^{\pm}(x)
\end{equation}
where $\tau^{\pm}(x)\in [-\infty,\infty]$. Here $\tau^{\pm}(x)\; =\; N^{\pm}(x)$ where $N^{\pm}(x)$ are given in $(\ref{eq:zero})$.
Whence,\\
\begin{equation}\label{eq:dynamics}
\hat{\alpha}_{\tau}\; f(x)[X^{+}_{p},X^{-}_{p};m]\; =\; f(x)[X^{+}_{p}\; +\; \tau^{+},X^{-}_{p}\; +\; \tau^{-};m]
\end{equation}
It can be shown that $\hat{\alpha}_{\tau}$ act as automorphisms on the algebra of observables \cite{bianca},\\
\begin{equation}
\hat{\alpha}_{\tau}\lbrace f(x)[X^{+}_{p},X^{-}_{p};m]\; ,\; f(x')[X^{+}_{p},X^{-}_{p};m]\rbrace\; =\; \lbrace \hat{\alpha}_{\tau}f(x)[X^{+}_{p},X^{-}_{p};m]\; ,\; \hat{\alpha}_{\tau}f(x')[X^{+}_{p},X^{-}_{p};m]\rbrace
\end{equation}
\hspace*{0.2in} So far it is not clear in what sense these transformations define temporal evolution (dynamics) of the complete observables. Easiest way to see this is by using an alternate characterization of the above automorphisms \cite{hajicek1}.\\
\hspace*{0.2in} Given a 1 parameter group of timelike diffeomorphisms $\theta(t)\; :\; M\rightarrow M$ of the auxiliary Minkowski background, one can associate to it a 1 parameter group of symplectic diffeomorphisms $\overline{\theta}(t)$ which are defined as follows.\\
\hspace*{0.2in} Given a point $m\; =\; (\phi,\; \pi_{\phi},\; X^{+},\; X^{-})$ in the constraint surface $\Gamma_{c}$ define $\overline{\theta}(t):\; \Gamma_{c}\rightarrow \Gamma_{c}$ as, $\overline{\theta}(t)(\phi,\; \pi_{\phi},\; X^{+},\; X^{-})\; =\; (\phi,\; \pi_{\phi},\; X^{+}\circ \theta(t),\; X^{-}\circ \theta(t))$. These symplectomorphisms shift the gauge-fixing slice $X\; =\; X_{p}$ to $X\circ \theta(t)\; =\; X_{p}$. This can also be understood as changing the prescribed embedding $X_{P}$ to some new embedding $\theta(-t)X_{p}$ \footnote{$\theta(-t)X_{p}$ is merely a notation. It is not to be understood in the same sense as $\theta(-t)X$}. Time evolution is the evolution of complete observables under above change.\\
\hspace*{0.2in} Consider for example 2 parameter family of timelike killing fields $V\; =\; A^{+}\partial_{+}\; +\; A^{-}\partial_{-}$ where $A^{\pm}$ are real numbers. using $\theta(t)\; =\; exp(tV)$\\
\begin{equation}
\begin{array}{lll}
X^{+}\rightarrow\; X^{+}\; +\; A^{+}t\\
\vspace*{0.2in}
X^{-}\rightarrow\; X^{-}\; +\; A^{-}t
\end{array}
\end{equation}
\hspace*{0.2in} The corresponding change in the gauge fixing slice $X\; =\; X_{p}$ is the same as that would be obtained by transforming $X^{\pm}_{p}$ as,\\
\begin{equation}\label{eq:three}
\begin{array}{lll}
X_{p}^{+}\rightarrow\; X_{p}^{+}\; -\; A^{+}t\\
\vspace*{0.2in}
X_{p}^{-}\rightarrow\; X_{p}^{-}\; -\; A^{-}t
\end{array}
\end{equation}
\hspace*{0.2in} As V is timelike, $\vert A^{+}\; +\; A^{-}\vert\; <\; \vert A^{+}\; -\; A^{-}\vert$. Comparing (\ref{eq:three}) with (\ref{eq:one}) we can rewrite (\ref{eq:one}) with $\tau^{+}\; =\; A^{+}t$ and $\tau^{-}\; =\; A^{-}t$. V is timelike implies,\\
\begin{equation}\label{eq:four}
\vert \tau^{+}\; +\; \tau^{-}\vert < \vert \tau^{+}\; -\; \tau^{-}\vert
\end{equation}
Thus  change in complete observables under (\ref{eq:one}) which satisfy the inequalities (\ref{eq:four}) define a class of time evolution for the system.\\
Whence by combining the complete set of Dirac observables obtained in the previous section with the notion of gauge-fixed slices in constraint surface we have defined dynamical observables of our theory.\\

\subsection{Canonical quantization of complete observables}

\hspace*{0.2in} Now we consider quantization of the complete observables $f(x)[X^{+}_{p},X^{-}_{p}]$ and $\pi_{f}(x)[X^{+}_{p},X^{-}_{p}]$ on $\mathcal{H}_{phy}$. As $X^{\pm}(x)$ are mere c-numbers, the canonical quantization of these observables follow directly from the quantization of the perennials ($a_{k},\; a_{k}*$) on $\mathcal{H}_{phy}$,\\
\begin{equation}
\begin{array}{lll}
(\widehat{f(x)[X^{+}_{p},X^{-}_{p};m]}_{can}\; \Psi)(f^{+}_{\overline{s}}\otimes f^{-}_{\overline{\overline{s}}})\; =\\
\vspace{0.1in}
\hspace*{1.3in} \Psi(\frac{1}{\sqrt{2\pi}}\int_{-\infty}^{+\infty}\frac{dk}{\vert k\vert}[\; \hat{a}_{k,T(\gamma(s))}e^{ik\cdot X_{p}(x)}\; +\; \hat{a}_{k,T(\gamma(s))}^{\dagger}e^{-ik\cdot X_{p}(x)}f^{+}_{\overline{s}}\otimes f^{-}_{\overline{\overline{s}}}])
\end{array}
\end{equation}
where $\Psi\in \mathcal{H}_{phy}$.\\
\emph{Note that although it is not possible to define an operator corresponding to f(x) on $\mathcal{H}_{kin}$ the corresponding complete observable is a well defined operator on $\mathcal{H}_{phy}$.}\\
One can similarly define an operator valued distribution for $\widehat{\pi_{f}(x)[X^{+}_{p},X^{-}_{p};m]}_{can}$ on $\mathcal{H}_{phy}$.\\
As before let $\Psi$ be a physical state given above.\\
\hspace*{0.2in} Recall that in order to define $\hat{a}_{k}$ , $\hat{a}^{\dagger}_{k}$ on $\mathcal{H}_{phy}$ we have to fix a pair of charge-networks $(\overline{s}_{0},\overline{\overline{s}}_{0})$ in the orbit of $(\overline{s},\overline{\overline{s}})$ and a pair of triangulations $(T(\overline{\gamma}_{0}(\overline{s}_{0})),T(\overline{\overline{\gamma}}_{0}(\overline{\overline{s}}_{0})))$.\\
\begin{equation}
\begin{array}{lll}
(\; \widehat{\pi_{f}(x)[X^{+}_{p},X^{-}_{p}]}_{can}\Psi\; )(f^{+}_{\overline{s}}\otimes f^{-}_{\overline{\overline{s}}})\; =\; \\
\\
\vspace*{0.2in}
\hspace*{2.7in} \Psi\; [\; \frac{i}{2\sqrt{\pi}}\int_{0}^{\infty}dk e^{ikX^{-}_{p}}\frac{(X^{-}_{p}(x)\; -\; X^{-}_{p}(v_{m}))}{\vert \triangle_{m}\vert} a_{k,T(\overline{\gamma}_{0})}\\
\\
\vspace*{0.1in}
\hspace*{2.7in} -\frac{i}{2\sqrt{\pi}}\int_{0}^{\infty}dk e^{-ikX^{-}_{p}}\frac{(X^{-}_{p}(x)\; -\; X^{-}_{p}(v_{m}))}{\vert \triangle_{m}\vert} a^{\dagger}_{k,T(\overline{\gamma}_{0})}\\
\\
\vspace*{0.1in}
\hspace*{2.7in} -\frac{i}{2\sqrt{\pi}}\int_{0}^{\infty}dk e^{ikX^{+}_{p}}\frac{(X^{+}_{p}(x)\; -\; X^{+}_{p}(v_{m}))}{\vert \triangle_{m}\vert} a_{k,T(\overline{\overline{\gamma}}_{0})}\\
\\
\vspace*{0.1in}
\hspace*{2.7in} +\frac{i}{2\sqrt{\pi}}\int_{0}^{\infty}dk e^{-ikX^{+}_{p}}\frac{(X^{+}_{p}(x)\; -\; X^{+}_{p}(v_{m}))}{\vert \triangle_{m}\vert} a^{\dagger}_{k,T(\overline{\overline{\gamma}}_{0})}\; ]\\
\vspace*{0.1in}
\hspace*{5.0in} (f^{+}_{\overline{s}_{0}}\otimes f^{-}_{\overline{\overline{s}}_{0}})
\end{array}
\end{equation}
\hspace*{0.2in} Here we have chosen $T(\overline{\gamma}_{0}(\overline{s}_{0}))$ and $T(\overline{\overline{\gamma}}_{0}(\overline{\overline{s}}_{0}))$ such that $\triangle_{m}$ is a simplex in both the triangulations and $v_{m}$, x are its initial and final vertices respectively.\\
\hspace*{0.2in} The Heisenberg dynamics is defined by promoting  automorphism of the algebra of complete observables (\ref{eq:dynamics}) to a automorphism on the algebra of corresponding operators, but this automorphism is not generated by any unitary operator. Thus the quantum dynamics is not unitary.\\
\hspace*{0.2in} There is a potential problem with the above definitions of complete observables in quantum theory. In classical theory although $X_{p}^{\pm}(x)$ are just parameters they are also the value of prescribed embeddings at spatial point x (recall the gauge fixing conditions $X^{\pm}\; =\; X^{\pm}_{p}$ required to define the complete observables). However in quantum theory there seems to be no relation between $X^{\pm}_{p}(x)$ and the embedding charges which label a given state. In view of this, we propose an alternative definition of $\widehat{f(x)[X^{+}_{p},X^{-}_{p}]}$, $\widehat{\pi_{f}(x)[X^{+}_{p},X^{-}_{p}]}$ in the next section and show how it leads to several interesting consequences in the quantum theory.\\
In this section we continue with the canonically quantized observables.
\\
\subsection{Complete observable from the dilaton field}

\hspace*{0.2in} We now apply the formalism we have developed so far to quantize the complete observable corresponding to the dilaton field on $\mathcal{H}_{phy}$. This operator is the starting point for the discussions about physical quantum geometry.\\
\hspace*{0.2in} Using the above expressions for the complete observables corresponding to the scalar field and its conjugate momenta, we can obtain an observable corresponding to the dilaton as follows.\\
\hspace*{0.2in} As shown in \cite{krv} the canonical transformation relating the dilaton to the embedding chart on the phase space is given by,\\
\begin{equation}
\begin{array}{lll}
y(x)\; =\; \lambda^{2}X^{+}(x)X^{-}(x)\; -\; \int_{\infty}^{x}d x_{1}X^{-'}(x_{1})\int_{\infty}^{x_{1}}d x_{2}\Pi_{-}(x_{2})\; +\; \int_{-\infty}^{x}d x_{1}X^{+'}(x_{1})\int_{-\infty}^{x_{1}}d x_{2}\Pi_{+}(x_{2})\\
\\
\vspace*{0.5in}
\hspace*{4.5in} +\; \int_{-\infty}^{\infty} X^{+}(x)\Pi_{+}(x)+\; \frac{m_{R}}{\lambda} .
\end{array}
\end{equation}
This expression can be rearranged using integration by parts as follows.\\
\begin{equation}
\begin{array}{lll}
y(x)\ =\ \lambda^{2}X^{+}(x)X^{-}(x)\ -\ X^{-}(x)\int_{\infty}^{x}d x_{1}\Pi_{-}(x_{1})\ +\ X^{+}(x)\int_{-\infty}^{x}d x_{1}\Pi_{+}(x_{1})\ +\\
\vspace*{0.1in}
\hspace*{0.6in} \int_{\infty}^{x}d x_{1}X^{-}(x_{1})\Pi_{-}(x_{1})\ +\ \int_{-\infty}^{x}d x_{1}X^{+}(x_{1})\Pi_{+}(x_{1})\ +\\
\vspace*{0.1in}
\hspace*{0.6in} \int_{-\infty}^{\infty} X^{+}(x)\Pi_{+}(x) +\; \frac{m_{R}}{\lambda} .
\end{array}
\end{equation}
\hspace*{0.2in} One can go to the constraint surface by solving for embedding momenta in terms of the scalar field and its conjugate momenta and by substituting the complete observable corresponding to the scalar field content, we obtain the observable corresponding to the dilaton. In the spacetime picture one can think of dilaton y(X), as a function of the spacetime scalar field, and the complete observable $y(x)[X^{+}_{p},X^{-}_{p}]$ (As mentioned above $x\in \Sigma$ should be thought of as a label set.) corresponds to the pull back of y(X) on a prescribed spatial slice when the free scalar field f(X) is pulled back on it.\\
Whence,\\
\begin{equation}
\begin{array}{lll}
y(x)[X^{+}_{p},X^{-}_{p}]\; =\; \lambda^{2}X^{+}_{p}(x)X^{-}_{p}(x)\; -\; \frac{X^{-}_{p}(x)}{4}\; \int_{\infty}^{x}d x_{1}\; \frac{Y_{-}(x_{1})[X^{+}_{p},X^{-}_{p}]^{2}}{X^{-'}_{P}(x_{1})}\\
\vspace*{0.2in}
-\;  \frac{X^{+}_{p}(x)}{4}\; \int_{-\infty}^{x}d x_{1}\; \frac{Y_{+}(x_{1})[X^{+}_{p},X^{-}_{p}]^{2}}{X^{+'}_{P}(x_{1})}\; +\; \int_{\infty}^{x}d x_{1}\; \frac{X^{-}_{p}(x_{1})}{X^{-}_{p'}(x_{1})}\; Y_{-}(x_{1})[X^{+}_{p},X^{-}_{p}]^{2}\\
\vspace*{0.2in}
+\; \int_{\infty}^{x}dx_{1}\; \frac{X^{+}_{p}(x_{1})}{X^{+}_{p'}(x_{1})}\; Y_{+}(x_{1})[X^{+}_{p},X^{-}_{p}]^{2} +\; \frac{m_{R}}{\lambda} .
\end{array}
\end{equation}
\hspace*{0.2in} We would like to promote $y(x)[X^{+}_{p},X^{-}_{p}]$ to an operator on $\mathcal{H}_{phy}$. This is the operator which contains the information about physical quantum geometry and thus is the most important ingredient in asking non-perturbative questions regarding fate of black-hole singularities, quantum fluctuation of event horizons  and even semi-classical issues like Hawking radiation.\\
\hspace*{0.2in} Once again it is important to note that as our previous complete observables, $\widehat{y(x)[X^{+}_{p},X^{-}_{p}]}_{can}$ is defined via its dual action on $\mathcal{H}_{kin}$. The derivation is given
in appendix A. Here we quote the final result.\\
\begin{equation}\label{eq:lame}
\begin{array}{lll}
\biggr[\; \widehat{y(x)[X^{+}_{p},X^{-}_{p}]}_{can}\Psi\; \biggl](\; f^{+}_{\overline{s}}\otimes f^{-}_{\overline{\overline{s}}}\otimes \vert m\rangle\; )= \\
\vspace*{0.1in}
\Psi\biggl(\ \biggl[ \lambda^{2}X_{p}^{+}(x)X_{p}^{-}(x)\ -\ X_{p}^{-}(x)\; \sum_{T_{x}(\overline{\gamma}_{0})}\bold{\hat{A}}\ -\ X_{p}^{+}(x)\; \sum_{\overline{T}_{x}(\overline{\overline{\gamma}}_{0})}\bold{\hat{B}}\\
\vspace*{0.1in}
+\ \sum_{\triangle_{m}\in T_{x}(\overline{\gamma_{0}})}X_{p}^{+}(v_{m})\bold{\hat{B}}\ +\ \sum_{\triangle_{m}\in T_{x}(\overline{\overline{\gamma_{0}})}}X_{p}^{-}(v_{m})\bold{\hat{A}}\ + \frac{\hat{m_{R}}}{\lambda}\ \biggr](\ f^{+}_{\overline{s}_{0}}\otimes\ f^{-}_{\overline{\overline{s}}_{0}})\ \biggr)
\end{array}
\end{equation}
where $\Psi\ =\ \vert\ N,\ N+1,\ {(k_{1}^{+},l_{1}^{+}),...,(k_{N}^{+},l_{N}^{+})}>_{+}\ \otimes\ \vert\ M,\ M+1,\ {(k_{1}^{-},l_{1}^{-}),...,(k_{N}^{-},l_{N}^{-})}>_{-}\otimes \vert m>$ is a physical state
and $f^{+}_{\overline{s}}\otimes f^{-}_{\overline{\overline{s}}}\otimes \vert m\rangle$ is an arbitrary charge network in $\mathcal{H}_{kin}$.
$T_{x}(\overline{\gamma_{0}})$ is the subcomplex of $(\overline{\gamma_{0}})$ from $-\infty$ to x.\\
\hspace*{0.2in} Thus we have a ``regulated'' expression $\widehat{y(x)[X_{p}^{+},X_{p}^{-}]}_{can}$ on $\mathcal{H}_{phy}$. Several ad-hoc choices are involved, notably fixing a pair of graphs $(\overline{\gamma_{0}},\overline{\overline{\gamma_{0}}})$ in the orbit of $(\overline{\gamma},\overline{\overline{\gamma}})$ and a pair of triangulations $(T(\overline{\gamma_{0}}),T(\overline{\overline{\gamma_{0}}}))$ corresponding to them.\\
\emph{It is interesting to note that we have an operator, and not an operator valued distribution for $\widehat{y(x)[X_{p}^{+},X_{p}^{-}]}_{can}$ on $\mathcal{H}_{phy}$.}\\
\hspace*{0.2in} In classical theory the singularities in physical spacetimes usually occur when $y(X)\; =\; 0$. Hence to understand the singularity structure of quantum geometry , the expectation value of $\widehat{y(x)[X_{p}^{+},X_{p}^{-}]}_{can}$ in a given physical state will play a central role. The expectation value of the dilaton is also a primary object in obtaining the semi-classical geometry from the non-perturbative quantum theory\cite{mikovic}. Thus in order to ask the physical questions using framework setup in this paper, evaluating the $\langle \Psi \vert \widehat{y(x)[X_{p}^{+},X_{p}^{-}]}_{can} \vert \Psi \rangle_{phy}$ is of crucial importance. Hence we now calculate the expectation value of $\widehat{y(x)[X^{+}_{p},X^{-}_{p}]}_{can}$ for a generic basis-state $\Psi$. It is straightforward to extend our results to obtain expectation value in an arbitrary state in $\mathcal{H}_{phy}$.\\
The calculations are summarized in appendix-B. The result is:\\
\hspace*{0.2in} for an arbitrary basis-state $\Psi$ in $\mathcal{H}_{phy}$. $\langle \Psi \vert \widehat{y(x)[X^{+}_{p},X^{-}_{p}]}_{can}\vert \Psi\rangle_{phy}$ equals,\\
\begin{equation}\label{eq:five}
\begin{array}{lll}
\langle \Psi \vert \widehat{y(x)[X^{+}_{p},X^{-}_{p}]}_{can}\vert \Psi\rangle_{phy}\; =\; \\
\vspace*{0.1in}
\Psi\biggr[\; \lambda^{2}X_{p}^{+}(x)X_{p}^{-}(x)\; + \frac{X_{p}^{-}(x)}{16\pi}\sum_{\triangle_{m}\in T_{x}}\mathcal{A}_{m}\; -\; \frac{X_{p}^{+}(x)}{16\pi}\sum_{\triangle_{m}\in T_{x}}\mathcal{B}_{m}\; -\; \frac{1}{16\pi}\sum_{\triangle_{m}\in T_{x}}X^{-}_{p}(v_{m})\mathcal{A}_{m}\\
\vspace*{0.1in}
-\; \frac{1}{16\pi}\sum_{\overline{T}_{x}}X^{+}_{p}(v_{m})\mathcal{B}_{m}\; +\; \frac{\hat{m}_{R}}{\lambda}\; \biggl](f^{+}_{\overline{s_{0}}}\otimes f^{-}_{\overline{\overline{s_{0}}}}\otimes \vert m>)
\end{array}
\end{equation}
Where $T_{x}$ is subcomplex from -$\infty$ to x and $\overline{T}_{x}$ is the subcomplex from x to $\infty$.\\
$\mathcal{A}_{m}$ and $\mathcal{B}_{m}$ are evaluated in appendix-B.\\
\begin{equation}
\begin{array}{lll}
\mathcal{A}_{m} & = & \frac{[X^{+}_{p}(v_{m})\; -\; X^{+}_{p}(v_{m-1})]^{2}}{X_{p}^{-}(v_{m})\; -\; X_{p}^{-}(v_{m-1})}\\
\\
\vspace*{0.2in}
& & \biggl\lbrace -4\sum_{n=0}^{N+1} P(\frac{1}{X^{+}_{p}(v_{m})\; -\; \frac{1}{2}(k^{+}_{n}+k^{+}_{n+1})})^{2}\; -\; 8\; P(\frac{1}{X_{p}^{+}(v_{m})})^{2}\biggr\rbrace
\end{array}
\end{equation}
\\
\begin{equation}
\begin{array}{lll}
\mathcal{B}_{m} & = & \frac{[X^{-}_{p}(v_{m})\; -\; X^{-}_{p}(v_{m-1})]^{2}}{X_{p}^{+}(v_{m})\; -\; X_{p}^{+}(v_{m-1})}\\
\\
\vspace*{0.2in}
& & \biggl\lbrace -4\sum_{n=0}^{N+1} P(\frac{1}{X^{-}_{p}(v_{m})\; -\; \frac{1}{2}(k^{-}_{n}+k^{-}_{n+1})})^{2}\; -\; 8\; P(\frac{1}{X_{p}^{-}(v_{m})})^{2}\biggr\rbrace
\end{array}
\end{equation}
\hspace*{0.2in} It might seem surprising that the matter charges $(l_{1},...l_{M})$ don't figure in these expressions (as they should!), but this is due to the fact that we have evaluated the expectation value in a basis-state. If $\Psi$ was an arbitrary state (linear combination of basis-states) $\langle \Psi \vert \widehat{y(x)[X^{+}_{p},X^{-}_{p}]}_{can}\vert \Psi\rangle_{phy}$
will depend on the matter charges as well.

\pagebreak

\subsection{Quantum complete observables}

\hspace*{0.2in} In this section we \emph{define} a set of observables in the quantum theory which we believe to be more appropriate counter-parts of the classical complete observables than the canonically quantized operators of the previous section.\\
\hspace*{0.2in} The basic idea is the following. Recall that the value of $f(x)[X^{+}_{p},X^{-}_{p}]$ on a gauge orbit $\mathcal{G}_{m}$ passing through  point m in the constraint surface is the value of the scalar field f(x) at $m^{'}$ where the gauge fixed slice $X^{\pm}\; =\; X^{\pm}_{p}$ intersects $\mathcal{G}_{m}$. Roughly speaking we try to mimic this construction directly at quantum level and obtain a new class of operators well defined on $\mathcal{H}_{phy}$.\\
The gauge fixing condition (at a point x in the spatial slice) translates on $\mathcal{H}_{kin}$ as,\\
\begin{equation}\label{eq:winter}
\hat{X}^{\pm}(x)f_{s}\; =\; X^{\pm}_{p}(x)f_{s}
\end{equation}
\hspace*{0.2in} As the quantum counter-part of gauge orbit in the classical theory is the orbit of charge network states in the quantum theory (represented by a state in $\mathcal{H}_{phy}$, we \cal{define} $\widehat{f(x)[X^{+}_{p},X^{-}_{p}]}$ as,\\
\begin{equation}\label{eq:jones}
\begin{array}{lll}
\widehat{f(x)[X^{+}_{p},X^{-}_{p};m]}\; \Psi\; :=\; \widehat{f(x)[X^{+}_{p},X^{-}_{p};m]}_{can}\; \Psi\\
\vspace*{0.1in}
\hspace*{2.4in} \mbox{if}\; X_{p}^{+}(x)\; \in\; (k_{1}^{+},...,k_{N}^{+})\; \mbox{and}\; X_{p}^{-}(x)\; \in\; (k_{1}^{-},...,k_{M}^{-}),\\
\vspace*{0.1in}
\hspace*{1.4in}:=\; 0\; \mbox{otherwise}
\end{array}
\end{equation}
where in the above,\\
$\Psi=$\\
$\vert N, N+1, (\theta^{+}_{1},...,\theta^{+}_{N}),[(k_{1}^{+},l_{1}^{+}),...,(k_{N}^{+},l_{N}^{+})]\rangle\otimes$ $\vert M, M+1, (\theta^{-}_{1},...,\theta^{-}_{M}), [(k_{1}^{-},l_{1}^{-}),...,(k_{M}^{-},l_{M}^{-})]\rangle$\\$\otimes
\vert m_{R}\rangle$ is a physical basis-state.\\
\hspace*{0.2in} This means the following. If the quantum gauge fixing condition (\ref{eq:winter}) intersects an orbit of equivalence class of states, then the action of $\widehat{f(x)[X^{+}_{p},X^{-}_{p};m]}$ on the orbit-$\Psi$ equals the action $\widehat{f(x)[X^{+}_{p},X^{-}_{p};m]}_{can}$. And those quantum orbits which do not intersect the gauge fixing slice (\ref{eq:winter}) are in the kernel of $\widehat{f(x)[X^{+}_{p},X^{-}_{p};m]}$.\\
\hspace*{0.2in} We would like to emphasize that  we are defining a two parameter $(X^{+}_{p}(x),X^{-}_{p}(x))$ family of observables in quantum theory and the claim is that they correspond to the complete observables $f(x)[X^{+}_{p},X^{-}_{p};m]$ of classical theory.  Eventually this has to be checked by semi-classical analysis but the motivation behind tying $X^{+}_{p}(x),X^{-}_{p}(x)$ to embedding charges $(k_{I}^{\pm})$ has been the interpretation of these complete observables.\\
Several features of this definition are worth noting.\\
1. Given a state $\Psi$, as $X_{p}^{\pm}(x)$ ``evolve'' from $-\infty$ to $\infty$, such that one of them increases and the other one decreases monotonically as given in \ref{eq:three} so as to match with the time evolution in classical theory,
$\widehat{f(x)[X^{+}_{p},X^{-}_{p};m]}\; \Psi\; \neq 0$ only when $X_{p}^{+}$ takes the discrete values $(\; k_{1}^{+},...,k_{N}^{+}\; )$ and $X_{p}^{-}$ takes the discrete values $(\; k_{1}^{-},...,k_{N}^{-}\; )$. In this sense the underlying dynamics in the quantum theory is discrete. It is important to note that we do not have a unitary dynamics so far. Automorphism of the algebra of complete observables is promoted to a automorphism on the algebra of corresponding operators, but this automorphism is not generated by any unitary operator.\\
2. Let us consider a (non-local) classical function $sin[k\cdot (X(x)\; -\; X(x'))]$ whose pull-back on prescribed slice
is given by $sin[k\cdot (X_{p}(x)\; -\; X_{p}(x'))]$. This function is trivially an observable and for fixed x, x' $X_{p}(x)$, $X_{p'}(x)$ are mere parameters. Thus one would naively expect to promote it to a multiple of identity operator on $\mathcal{H}_{phy}$. However from our definition of complete observable in quantum theory it turns out that the correct quantization of $sin[k\cdot (X_{p}(x)\; -\; X_{p}(x'))]$ is given by,\\
\begin{equation}
(\widehat{sin[k\cdot (X_{p}(x)\; -\; X_{p}(x'))]}\; \Psi)(f_{\gamma})\; =\; \Psi(sin[k\cdot (X_{p}(x)\; -\; X_{p}(x'))]\bold{1}f_{\gamma})
\end{equation}
if $X^{\pm}(x)\; \in\; (k_{1}^{\pm},...,k_{N}^{\pm})$ and $X^{\pm}(x')\; \in\; (k_{1}^{\pm},...,k_{N}^{\pm})$
and zero otherwise.\\
\hspace*{0.2in} This observation ensures that if as mentioned earlier, there exist class of semi-classical states on which the commutator algebra generated by $(a_{k}, a_{k}^{\dagger})$ mirrors the classical Poisson algebra (to leading order in $\hbar$) then on those states one can easily show that the commutator between $f(x)[X^{+}_{p},X^{-}_{p}]$ , $f(x')[X^{+}_{p},X^{-}_{p}]$ vanishes (again to leading order in $\hbar$) iff $\vec{X_{p}}(x)$ and $\vec{X_{p}}(x')$ are spacelike separated.\\
One can similarly define an operator valued distribution for $\widehat{\pi_{f}(x)[X^{+}_{p},X^{-}_{p};m]}$ on $\mathcal{H}_{phy}$.\\
As before, let $\Psi$ be a physical state given above. Then,\\
\begin{equation}\label{eq:midget}
\widehat{\pi_{f}(x)[X^{+}_{p},X^{-}_{p}]}\; \Psi\; :=\; \widehat{\pi_{f}(x)[X^{+}_{p},X^{-}_{p}]}_{can}\; \Psi
\end{equation}
if $(X^{+}_{p}(x),X^{+}_{p}(v_{m}))\in (k^{+}_{I},k^{+}_{I+1}), 1\leq I\leq N$ and $(X^{-}_{p}(x),X^{-}_{p}(v_{m}))\in (k^{-}_{J},k^{-}_{J+1}), 1\leq J\leq M$.\\
and,\\
\begin{equation}
\widehat{\pi_{f}(x)[X^{+}_{p},X^{-}_{p}]}\; \Psi\; :=\; 0
\end{equation}
otherwise.\\
\hspace*{0.2in} We have chosen $T(\overline{\gamma}_{0})$ and $T(\overline{\overline{\gamma}}_{0})$ such that $\triangle_{m}$ is a simplex in both the triangulations and $v_{m}$, x
are its initial and final vertices respectively.\\

\subsection{New dilaton operator}

\hspace*{0.2in} Using the quantum observables $\widehat{f(x)[X^{+}_{p},X^{-}_{p}]}$ , $\widehat{\pi_{f}(x)[X^{+}_{p},X^{-}_{p}]}$ one can define a new complete observable corresponding to the dilaton ( $\widehat{y(x)[X^{+}_{p},X^{-}_{p}]}$ ) in the quantum theory. This operator will obviously differ from $\widehat{y(x)[X^{+}_{p},X^{-}_{p}]}_{can}$ and if as argued earlier $\widehat{f(x)[X^{+}_{p},X^{-}_{p}]}$ , $\widehat{\pi_{f}(x)[X^{+}_{p},X^{-}_{p}]}$ are the appropriate dynamical observables of the quantum theory then information about quantum geometry will be encoded in $\widehat{y(x)[X^{+}_{p},X^{-}_{p}]}$.\\
\hspace*{0.2in} Using defining equations (\ref{eq:jones}) , (\ref{eq:midget}) for $\widehat{f(x)[X^{+}_{p},X^{-}_{p}]}$ and $\widehat{\pi_{f}(x)[X^{+}_{p},X^{-}_{p}]}$ we can show that the formal expression for $\widehat{y(x)[X^{+}_{p},X^{-}_{p}]}$
remains same as that for $\widehat{y(x)[X^{+}_{p},X^{-}_{p}]}_{can}$ (\ref{eq:lame}) with the proviso that only the following terms contribute.\\
1. First term is non-zero iff $X^{+}_{p}(x)\in (k_{1}^{+},...,k_{N}^{+},\frac{k_{1}^{+}+k_{2}^{+}}{2},...,\frac{k_{N-1}^{+}+k_{N}^{+}}{2})$ and \\
$X^{-}_{p}(x)\in (k_{1}^{-},...,k_{M}^{-},\frac{k_{1}^{-}+k_{2}^{-}}{2},...,\frac{k_{M-1}^{-}+k_{M}^{-}}{2})$.\\

2. Second term ($X_{p}^{-}(x)\; \sum_{T_{x}(\overline{\gamma}_{0})}\bold{\hat{A}}$) is non-zero iff $X^{+}_{p}(x)\in (k_{1}^{+},...,k_{N}^{+})$\\
a. There are various summations involved in this expression. In the outermost summation $\sum_{\triangle_{m}}$
only those simplices $\triangle_{m}$ contribute for which\\
\begin{displaymath}
(X_{p}^{+}(v_{m+1}),X_{p}^{+}(V_{m}))\in [(k_{I+1}^{+},k_{I}^{+})\; or\; (\frac{k_{I+1}^{+}+k_{I}^{+}}{2},k_{I})\; or\; (k_{I+1}^{+},\frac{k_{I+1}^{+}+k_{I}^{+}}{2})]
\end{displaymath}
for some $I\in (1,...,N-1)$.\\
\begin{displaymath}
(X_{p}^{-}(v_{m+1}),X_{p}^{-}(V_{m}))\in [(k_{I+1}^{-},k_{I}^{-})\; or\; (\frac{k_{I+1}^{-}+k_{I}^{-}}{2},k_{I}^{-})\; or\; (k_{I+1}^{-},\frac{k_{I+1}^{-}+k_{I}^{-}}{2})]
\end{displaymath}
for some $I\in (1,...,M-1)$.\\
\hspace*{0.2in} One immediate consequence of this criterion is that, as we require $X_{p}^{\pm'}(x)\neq 0$, \emph{only} a finite number of simplices will contribute to the outermost summation $\sum_{\triangle_{m}}$.\\
3. Third term ( $X_{p}^{+}(x)\; \sum_{\overline{T}_{x}(\overline{\overline{\gamma}}_{0})}\bold{\hat{B}}$ ) will be non-vanishing iff\\
$X^{-}_{p}(x)\in (k_{1}^{-},...,k_{M}^{-},\frac{k_{1}^{-}+k_{2}^{-}}{2},...,\frac{k_{M-1}^{-}+k_{M}^{-}}{2})$. Only those simplices in the Riemann sum $\sum_{\overline{T}_{x}(\overline{\overline{\gamma}}_{0})}$ which satisfy condition 2-a.\\
Similar remarks apply to fourth and fifth terms.\\
Let us summarize.\\
Action of $\widehat{y(x)[X^{+}_{p},X^{-}_{p}]}$ on $\mathcal{H}_{phy}$ can be defined by its dual action as,\\
\begin{equation}\label{eq:dino}
\begin{array}{lll}
\biggr[\; \widehat{y(x)[X^{+}_{p},X^{-}_{p}]}^{'}\ \vert\Psi\ \rangle\; \biggl](\; f^{+}_{\overline{s}}\otimes f^{-}_{\overline{\overline{s}}}\otimes \vert m\rangle\; )= \\
\vspace*{0.1in}
\langle\ \Psi \vert\ \biggl(\ \biggl[ \lambda^{2}X_{p}^{+}(x)X_{p}^{-}(x)\ -\ X_{p}^{-}(x)\; \sum_{T_{x}(\overline{\gamma}_{0})}\bold{\hat{A}}\ -\ X_{p}^{+}(x)\; \sum_{\overline{T}_{x}(\overline{\overline{\gamma}}_{0})}\bold{\hat{B}}\\
\vspace*{0.1in}
+\; \sum_{\triangle_{m}\in T_{x}(\overline{\gamma_{0}})}X_{p}^{+}(v_{m})\bold{\hat{B}}\ +\ \sum_{\triangle_{m}\in T_{x}(\overline{\overline{\gamma_{0}})}}X_{p}^{-}(v_{m})\bold{\hat{A}}\ + \frac{\hat{m_{R}}}{\lambda}\ \biggr](\ f^{+}_{\overline{\gamma}_{0}}\otimes\ f^{-}_{\overline{\overline{\gamma}}_{0}})\ \biggr)
\end{array}
\end{equation}
1. The first 3 terms in (\ref{eq:dino}) are non-zero iff $(X^{+}_{p}(x)\in (k_{1}^{+},...,k_{N}^{+},\frac{k_{1}^{+}+k_{2}^{+}}{2},...,\frac{k_{N-1}^{+}+k_{N}^{+}}{2})$ and $(X^{-}_{p}(x)\in (k_{1}^{-},...,k_{M}^{-},\frac{k_{1}^{-}+k_{2}^{-}}{2},...,\frac{k_{M-1}^{-}+k_{M}^{-}}{2})$.\\
2.There are various Riemann sums involved in each term involving $\bold{\hat{A}}$, $\bold{\hat{B}}$. In the summations over
$T_{x}$ in \ref{eq:dino}, only those simplices contribute for which $(X_{p}^{\pm}(v_{m}),X_{p}^{\pm}(v_{m+1}))$ lie in a certain finite set as explained above.\\

\subsection{Zeros of dilaton operator}

\hspace*{0.2in} As a naive application of the above expression for expectation value of $\widehat{y(x)[X^{+}_{p},X^{-}_{p}]}$ we show the existence of a set of states which do not lie in the kernel of $\widehat{y(x)[X^{+}_{p},X^{-}_{p}]}$ but with respect to which $\langle \widehat{y(x)[X^{+}_{p},X^{-}_{p}]} \rangle$ is zero. We do not claim that these states shed any light on the singularity structure of quantum spacetime for which defining $\hat{\frac{1}{y}}$ operator is essential. We only give this example to show that if instead of $\widehat{y(x)[X^{+}_{p},X^{-}_{p}]}_{can}$ we take $\widehat{y(x)[X^{+}_{p},X^{-}_{p}]}$ as the defining operator for quantum geometry, then it is possible to hunt for states in which the inverse physical metric $g^{\mu\nu}\; =\; y\gamma^{\mu\nu}$ is degenerate.\\
\hspace*{0.2in} Given a point x and prescribed embeddings $X^{\pm}_{p}:\Sigma \rightarrow M$ one can always ``construct'' a (large) set of states in which all the terms in ($\ref{eq:five}$) except $\sum_{\overline{T}_{x}}X^{+}_{p}(v_{m})\mathcal{B}_{m}$ contribute and terms $\frac{X_{p}^{-}(x)}{16\pi}\sum_{\triangle_{m}\in T_{x}}\mathcal{A}_{m}$, $\frac{1}{16\pi}\sum_{\triangle_{m}\in T_{x}}X^{-}_{p}(v_{m})\mathcal{A}_{m}$ cancel each other out.\\
\hspace*{0.2in} This can be done as follows. choose a state $\Psi$ and corresponding $(\overline{s}_{0}, \overline{\overline{s}}_{0})$ such that  x is one of the vertices of both the graphs and say the nearest vertex to the left be denoted by $v_{m-1}$. Also Let the embedding charges in $\Psi$ be such that $(X^{+}_{p}(x),X^{+}_{p}(v_{m-1})\; =\; (k^{+}_{I},k^{+}_{I-1})$ for some I and
$(X^{-}_{p}(x),X^{-}_{p}(v_{m-1})\; =\; (k^{-}_{J},k^{-}_{J-1})$ for some J, but for no other vertex of the triangulation $X^{\pm}_{v_{k}}\in (k^{\pm}_{1},...,k^{pm}_{N})$. Now it is easy to show that if both $X^{\pm}(x)$ are positive and if $X_{p}^{+}(v_{m-1}) < X^{+}_{p}(x)$ then it is possible to solve for a positive value of m such that $\langle \widehat{y(x)[X^{+}_{p},X^{-}_{p}]} \rangle\; =\; 0$.\\
\hspace*{0.2in} One way to understand singularity structure in quantum theory would be to check the boundedness of $\hat{\frac{1}{y}}$ operator at those values of $(X^{+}_{p}(x), X^{-}_{p}(x))$ at which the above zero occurs. Also If for a range of $(X^{+}_{p}(x), X^{-}_{p}(x))$, $\langle \widehat{y(x)[X^{+}_{p},X^{-}_{p}]} \rangle$ becomes negative (implying a change in the signature of the physical metric) then this opens up the possibility of evolving the quantum geometry through the classical singularity \cite{krv}. As we do not know how to define $\hat{\frac{1}{y}}$ on $\mathcal{H}_{phy}$, these issues remain open.

\section{Discussion}

\hspace*{0.2in} The primary aim of this work has been to obtain a quantum theory of dilaton gravity by combining the ideas of parametrized field theory and polymer (loop) quantization. We started with a parametrized field theory which is canonically equivalent to the KRV action. By choosing appropriate quantum algebras for the embedding and matter sectors, we obtained a Hilbert space which carries a unitary (and anomaly-free) representation of the space-time diffeomorphism group. Using the so called group averaging method, we were able to get rid of the quantum gauge degrees of freedom and obtain the physical spectrum of the theory in a rather straightforward manner. The parametrized field theory framework gave us a complete set of Dirac observables which we could promote to well defined operators on $\mathcal{H}_{phy}$. This required ad hoc choices of triangulations and the final operators are dependent on choice of triangulation. This ad-hocness permeates all the consequent constructions and calculations. We hope that eventually a better (regularization-independent) scheme emerges to quantize such observables or at least regularization dependent scheme introduced in this paper can find more physical justification.\\
\hspace*{0.2in} Unlike the Fock space which by definition is an irreducible representation of the  Poisson algebra of mode  oscillators ($a_{k}\; ,\; a^{*}_{k}$), $\mathcal{H}_{phy}$ carries a representation of a deformed algebra. It is a faithful deformation of the classical algebra in the sense that all the corrections are $O(\hbar)$. It is an interesting open question to hunt for the full quantum algebra  and try to find physical interpretation of its elements which do not have a well defined classical limit (the commutator $[\hat{a}_{k},\hat{a}_{l}]$ defines one such element).\\
\hspace*{0.2in} Time evolution could be defined in this model by using the complete (dynamical) observables and seeing how they change under symplectomorphisms which arise from certain diffeomorphisms of the background spacetime. We gave two inequivalent definitions of complete observables in quantum theory and the corresponding Heisenberg dynamics. The first definition was through the canonical quantization of classical observables. However as argued in the paper, we believe that canonically quantizing the complete observables does not preserve its physical interpretation. This led us to the second definition which captures the relational nature of classical complete observables in a more transparent manner than the canonically quantized counterparts. Adapting the second definition to define dynamical observables in the quantum theory implies discrete temporal evolution. Finally using either definition of quantum complete observables, we defined physical dilaton operator which is well defined without smearing on $\mathcal{H}_{phy}$. We also calculated its expectation value on an arbitrary basis-state in $\mathcal{H}_{phy}$ (for a specific choice of triangulation). This, we believe, gives us a framework to address the semi-classical and non-perturbative issues arising in the CGHS model.\\
\hspace*{0.2in} We would once again like to emphasize that the most worrisome feature of our work is the regularization dependence of observables. It is imperative that one finds a physical interpretation for this dependence or give a more sophisticated (regularization independent) quantization scheme.\\
\hspace*{0.2in} The most immediate open questions we need to answer are related to semi-classical analysis and inverse dilaton operator. We need to define physical semi-classical states and show that classical physics is indeed captured by expectation values. If the expectation values of $\widehat{f(x)[X^{+}_{p},X^{-}_{p}]}$ , $\widehat{\pi_{f}(x)[X^{+}_{p},X^{-}_{p}]}$ in the semi-classical states equals $f(x)[X^{+}_{p},X^{-}_{p}]$ and $\pi_{f}(x)[X^{+}_{p},X^{-}_{p}]$ to zeroth order in $\hbar$ then their use as dynamical observables in quantum theory is justified. Finally in order to prove the folklore that classical Black hole singularity is resolved in the quantum theory, defining $\hat{\frac{1}{y}}$ on $\mathcal{H}_{phy}$ is essential. We plan to pursue some of these questions in the near future.\\

{\bf Acknowledgements}: I am grateful to Ghanshyam Date for numerous discussions, for careful reading of the manuscript and for his support. I thank Bobby Ezhuthachan and Bala Sathiapalan for stimulating discussions regarding \cite{g-s}, \cite{jackiw1}. Finally I am indebted to Madhavan Varadarajan for numerous discussions, his critique of some of the results,  suggestions on the manuscript and most importantly for his friendship.

\appendix

\newcommand{\appsection}[1]{\let\oldthesection\thesection
  \renewcommand{\thesection}{Appendix \oldthesection}
  \section{#1}\let\thesection\oldthesection}

\section{Definition of the physical dilaton operator}

\hspace*{0.2in} In this appendix we canonically quantize the complete observable corresponding to the dilaton. We start with an arbitrary basis state $\Psi$ and derive the (dual) action of $\widehat{y(x)[X^{+}_{p},X^{-}_{p}]_{can}}$ on it. The operator can be extended to an arbitrary physical state by linearity.
\\
\hspace*{0.2in} Let $\Psi\ =\ \vert\ N,\ N+1,\ {(k_{1}^{+},l_{1}^{+}),...,(k_{N}^{+},l_{N}^{+})}>_{+}\ \otimes\ \vert\ M,\ M+1,\ {(k_{1}^{-},l_{1}^{-}),...,(k_{N}^{-},l_{N}^{-})}>_{-}\otimes \vert m>$. Without loss of generality, let us assume that N $>$ M.
\hspace*{0.2in} Also let $f^{+}_{\overline{s}}\otimes\ f^{-}_{\overline{\overline{s}}}\otimes \vert m>$ be any state in $\mathcal{H}_{kin}$. In the orbit of $(\overline{s},\; \overline{\overline{s}})$ fix a pair $(\overline{s}_{0},\;  \overline{\overline{s}}_{0})$, with the corresponding graphs $(\overline{\gamma_{0}}(\overline{s}_{0}),\; \overline{\overline{\gamma_{0}}}( \overline{\overline{s}}_{0}))$.
\hspace*{0.2in} Using our scheme of how to define operator corresponding to complete observable in quantum theory $\widehat{y(x)[X^{+}_{p},X^{-}_{p}]'}$ is as follows.
\begin{equation}
\begin{array}{lll}
[\widehat{y(x)[X^{+}_{p},X^{-}_{p}]}_{can'}\Psi\; ](f^{+}_{\overline{s}}\otimes\ f^{-}_{\overline{\overline{s}}}\otimes \vert m>)\; := \\
\biggl[\; [\lambda^{2}X^{+}_{p}(x)X^{-}_{p}(x)\bold{1}\; -\; \frac{X^{-}_{p}(x)}{4} \int_{\infty}^{x}d\overline{x}\;  \frac{\widehat{Y_{-}(\overline{x})[X^{+}_{p},X^{-}_{p}]^{2}}}{X_{P}^{-'}(\overline{x})}\; -\; \frac{X^{+}_{p}(x)}{4} \int_{-\infty}^{x}d\overline{x}\; \frac{\widehat{Y_{+}(\overline{x})[X^{+}_{p},X^{-}_{p}]^{2}}}{X_{P}^{+'}(\overline{x})}\\
\vspace*{0.1in}
+\; \int_{\infty}^{x}d\overline{x}\; \frac{X^{-}_{p}(\overline{x})}{X^{-}_{p'}(\overline{x})}\ \widehat{Y_{-}(\overline{x})[X^{+}_{p},X^{-}_{p}]^{2}}\; +\;  \int_{\infty}^{x}d\overline{x}\; \frac{X^{+}_{p}(\overline{x})}{X^{+}_{p'}(\overline{x})}\; \widehat{Y_{+}(\overline{x})[X^{+}_{p},X^{-}_{p}]^{2}}\; +\;  \frac{\hat{m_{R}}}{\lambda}\; ]\Psi \biggr]\\
\vspace*{0.1in}
\hspace*{4.0in}(f^{+}_{\overline{\gamma_{o}}}\otimes\ f^{-}_{\overline{\overline{s_{0}}}}\; \otimes \vert m>)
\end{array}
\end{equation}
Where the first term is a multiple of identity operator. Note that as x is a fixed point in the spatial slice, $X^{\pm}_{p}(x)$ are just parameters.\\
\hspace*{0.2in} Remaining terms can be calculated using expressions for $f(x)[X^{+}_{p},X^{-}_{p}]$, $\pi_{f}(x)[X^{+}_{p},X^{-}_{p}]$ on $\mathcal{H}_{kin}$.\\
\underline{Term 2}\\
Second term in $\widehat{y(x)[X^{+}_{p},X^{-}_{p}]'}$ is given by,\\
\begin{equation}
\begin{array}{lll}
\Psi(\; \frac{X^{-}_{p}(x)}{4}\widehat{\int_{\infty}^{x}d\overline{x}\ Y_{-}(\overline{x})^{\dagger}[X^{+}_{p},X^{-}_{p}]^{2}}\ f^{+}_{\overline{s}}\otimes\ f^{-}_{\overline{\overline{s}}}\; )\\
\vspace*{0.2in}
= \Psi\frac{1}{16\pi}\sum_{\triangle_{m}\in T_{x}(\overline{\gamma_{0}})}\ (X_{p}^{+}(v_{m+1})\ -\ X_{p}^{+}(v_{m}))^{2}((X_{p}^{-}(v_{m+1})\ -\ X_{p}^{-}(v_{m}))^{-1}\\
\vspace*{0.1in}
\hspace*{0.3in} \biggr[\; {\int_{-\infty}^{0}dke^{ikX_{p}^{+}(v_{m})} \sum_{\triangle_{n}}e^{-ik\hat{X}^{+}(v_{n})}[h_{\triangle_{n}}(Y^{+})\ -\ h_{\triangle_{n}^{-1}}(Y^{+})]\ }\times\\
\vspace*{0.1in}
\hspace*{0.4in} {\int_{-\infty}^{0}d\overline{k}e^{i\overline{k}X_{p}^{+}(v_{m})} \sum_{\triangle_{l}}e^{-i\overline{k}\hat{X}^{+}(v_{l})}[h_{\triangle_{l}}(Y^{+})\ -\ h_{\triangle_{l}^{-1}}(Y^{+})]\ }\\
\vspace*{0.1in}
\hspace*{0.3in} -{\int_{-\infty}^{0}dke^{-ikX_{p}^{+}(v_{m})} \sum_{\triangle_{n}}e^{ik\hat{X}^{+}(v_{n})}[h_{\triangle_{n}}(Y^{+})\ -\ h_{\triangle_{n}^{-1}}(Y^{+})]\ }\times\\
\vspace*{0.1in}
\hspace*{0.4in} {\int_{-\infty}^{0}d\overline{k}e^{i\overline{k}X_{p}^{+}(v_{m})} \sum_{\triangle_{l}}e^{-i\overline{k}\hat{X}^{+}(v_{l})}[h_{\triangle_{l}}(Y^{+})\ -\ h_{\triangle_{l}^{-1}}(Y^{+})]\ }\\
\vspace*{0.1in}
\hspace*{0.3in} -{\int_{-\infty}^{0}dke^{ikX_{p}^{+}(v_{m})}\ \sum_{\triangle_{n}}e^{-ik\hat{X}^{+}(v_{n})}[h_{\triangle_{n}}(Y^{+})\ -\ h_{\triangle_{n}^{-1}}(Y^{+})]}\times\\
\vspace*{0.1in}
\hspace*{0.4in} {\int_{-\infty}^{0}dke^{ikX_{p}^{+}(v_{m})}\ \sum_{\triangle_{n}}e^{-ik\hat{X}^{+}(v_{n})}[h_{\triangle_{n}}(Y^{+})\ -\ h_{\triangle_{n}^{-1}}(Y^{+})]}\\
\vspace*{0.1in}
\hspace*{0.3in} +{\int_{-\infty}^{0}dke^{-ikX_{p}^{+}(v_{m})}\ \sum_{\triangle_{n}}e^{ik\hat{X}^{+}(v_{n})}[h_{\triangle_{n}}(Y^{+})\ -\ h_{\triangle_{n}^{-1}}(Y^{+})]}\times\\
\vspace*{0.1in}
\hspace*{0.4in} {\int_{-\infty}^{0}d\overline{k}e^{-i\overline{k}X_{p}^{+}(v_{m})}\ \sum_{\triangle_{l}}e^{i\overline{k}\hat{X}^{+}(v_{l})}[h_{\triangle_{l}}(Y^{+})\ -\ h_{\triangle_{l}^{-1}}(Y^{+})]}\;  \biggl]\; (f^{+}_{\overline{s_{0}}}\otimes\ f^{-}_{\overline{\overline{s_{0}}}})
\end{array}
\end{equation}
\\
where $T(\overline{\gamma_{0}})$ is a fixed triangulation adapted to $\overline{\gamma_{0}}$.\footnote{In this section we denote $\overline{\gamma_{0}}$ by $\overline{\gamma_{0}}(\overline{s_{0}})$ to avoid proliferation of symbols.}
 Here adapted means in the image of the graph all the vertices of the graph are vertices of triangulation and ``outside'' the graph $T(\overline{\gamma_{0}})$ is arbitrary( However we will make a more specific choice of triangulation when calculating expectation value. It is the same choice that we made when evaluating the commutators between perennials.) As it should be clear from the classical expression for $y(x)[X^{+}_{p},X^{-}_{p}]$, $T_{x}(\overline{\gamma_{0}})$ is the triangulation of the spatial manifold from x to $\infty$.\\
In the above expression we will denote everything inside ${\bf [}$..${\bf ]}$ as ${\bf \hat{A}}$.\\
\underline{Term 3}\\
\hspace*{0.2in} Here we denote the triangulation adapted to $\overline{\overline{\gamma}}_{0}$ $T(\overline{\overline{\gamma}}_{0})$ and the sub-complex ranging from $-\infty$ to x as $\overline{T}_{x}(\overline{\overline{\gamma}}_{0})$.\\
\begin{equation}
\begin{array}{lll}
[\; \frac{X^{+}_{p}(x)}{4} \widehat{\int_{-\infty}^{x}d\overline{x}\ \frac{Y_{+}(\overline{x})[X^{+}_{p},X^{-}_{p}]^{2}}{X_{p}^{+'}}(\overline{x})'}\Psi\; ](f^{+}_{\overline{s}}\otimes\ f^{-}_{\overline{\overline{s}}}\; )\\
\vspace*{0.2in}
=\; \frac{1}{16\pi}\Psi\; \sum_{\triangle_{m}\in \overline{T}_{x}(\overline{\overline{\gamma_{0}}})}\ (X_{p}^{-}(v_{m+1})\ -\ X_{p}^{-}(v_{m}))^{2}((X_{p}^{+}(v_{m+1})\ -\ X_{p}^{+}(v_{m}))^{-1})\\
\vspace*{0.1in}
\hspace*{0.3in} \biggr[\; {\int_{0}^{\infty}dke^{ikX_{p}^{-}(v_{m})} \sum_{\triangle_{n}\in T(\overline{\overline{\gamma_{0}}})}e^{-ik\hat{X}^{-}(v_{n})}[h_{\triangle_{n}}(Y^{-})\ -\ h_{\triangle_{n}^{-1}}(Y^{-})]\ }\times\\
\vspace*{0.1in}
{\hspace*{0.4in} \int_{0}^{\infty}d\overline{k}e^{i\overline{k}X_{p}^{-}(v_{m})} \sum_{\triangle_{l}\in T(\overline{\overline{\gamma_{0}}})}e^{-i\overline{k}\hat{X}^{-}(v_{l})}[h_{\triangle_{l}}(Y^{-})\ -\ h_{\triangle_{l}^{-1}}(Y^{-})]\; }\\
\vspace*{0.1in}
\hspace*{0.3in} -{\int_{0}^{\infty}dke^{ikX_{p}^{-}(v_{m})} \sum_{\triangle_{n}\in T(\overline{\overline{\gamma_{0}}})}e^{-ik\hat{X}^{-}(v_{n})}[h_{\triangle_{n}}(Y^{-})\ -\ h_{\triangle_{n}^{-1}}(Y^{-})]\ }\times\\
\vspace*{0.1in}
\hspace*{0.4in} {\int_{0}^{\infty}d\overline{k}e^{i\overline{k}X_{p}^{-}(v_{m})} \sum_{\triangle_{l}}e^{-i\overline{k}\hat{X}^{-}(v_{l})}[h_{\triangle_{l}}(Y^{-})\ -\ h_{\triangle_{l}^{-1}}(Y^{-})]\ }\\
\vspace*{0.1in}
\hspace*{0.3in} -{\int_{0}^{\infty}dke^{-ikX_{p}^{-}(v_{m})}\ \sum_{\triangle_{n}}e^{ik\hat{X}^{-}(v_{n})}[h_{\triangle_{n}}(Y^{-})\ -\ h_{\triangle_{n}^{-1}}(Y^{-})]}\times\\
\vspace*{0.1in}
\hspace*{0.4in} {\int_{0}^{\infty}d\overline{k}e^{i\overline{k}X_{p}^{-}(v_{m})}\ \sum_{\triangle_{l}}e^{-i\overline{k}\hat{X}^{-}(v_{l})}[h_{\triangle_{l}}(Y^{-})\ -\ h_{\triangle_{l}^{-1}}(Y^{-})]}\\
\vspace*{0.1in}
\hspace*{0.3in} +\; {\int_{0}^{\infty}dke^{-ikX_{p}^{-}(v_{m})}\ \sum_{\triangle_{n}}e^{-ik\hat{X}^{-}(v_{n})}[h_{\triangle_{n}}(Y^{-})\ -\ h_{\triangle_{n}^{-1}}(Y^{-})]}\\
\vspace*{0.1in}
\hspace*{0.4in} {\int_{0}^{\infty}d\overline{k}e^{-i\overline{k}X_{p}^{-}(v_{m})}\ \sum_{\triangle_{l}}e^{-i\overline{k}\hat{X}^{-}(v_{l})}[h_{\triangle_{l}}(Y^{-})\ -\ h_{\triangle_{l}^{-1}}(Y^{-})]}\;  \biggl]f^{+}_{\overline{s_{0}}}\otimes\ f^{-}_{\overline{\overline{s_{0}}}}
\end{array}
\end{equation}
We denote everything inside $\sum_{\overline{T}_{x}(\overline{\overline{\gamma_{0}}})}\; as \bold{\hat{B}}$.\\
\underline{Term 4}\\
\\
This term can be directly derived from term-2.
\\
\begin{equation}
\begin{array}{lll}
\Psi(\frac{1}{4}\int_{\infty}^{x}d\overline{x}\ X_{p}^{-}(\overline{x})\frac{\widehat{Y_{-}(\overline{x})^{\dagger}[X^{+}_{p},X^{-}_{p}]^{2}}}{X_{p}^{-'}(\overline{x})}\ (f^{+}_{\overline{s}}\otimes\ f^{-}_{\overline{\overline{s}}})\ )\\
\vspace*{0.1in}
\\
=\ \Psi(\; \sum_{\triangle_{m}\in T_{x}(\overline{\gamma_{0}})}X_{p}^{-}(v_{m})\bold{\hat{A}}(f^{+}_{\overline{s}}\otimes\ f^{-}_{\overline{\overline{s}}})\; )
\end{array}
\end{equation}
\\
\underline{Term 5}\\
Finally using term-3 we see that,\\
\begin{equation}
\begin{array}{lll}
\Psi(\frac{1}{4}\int_{\infty}^{x}d\overline{x}\ X_{p}^{+}(\overline{x})\frac{\widehat{Y_{+}(\overline{x})^{\dagger}[X^{+}_{p},X^{-}_{p}]^{2}}}{X_{p}^{+'}(\overline{x})}\ (f^{+}_{\overline{s}}\otimes\ f^{-}_{\overline{\overline{s}}})\ )\\
\vspace*{0.1in}
\\
=\ \Psi(\; \sum_{\triangle_{m}\in T_{x}(\overline{\gamma_{0}})}X_{p}^{+}(v_{m})\bold{\hat{B}}(f^{+}_{\overline{s}}\otimes\ f^{-}_{\overline{\overline{s}}})\; )
\end{array}
\end{equation}
Whence the final expression for the dilaton operator $\widehat{y(x)[X^{+}_{p},X^{-}_{p}]}$ at a given point x is,\\
\begin{equation}\label{eq:lame}
\begin{array}{lll}
\biggr[\; \widehat{y(x)[X^{+}_{p},X^{-}_{p}]}_{can'}\Psi\; \biggl](\; f^{+}_{\overline{s}}\otimes f^{-}_{\overline{\overline{s}}}\otimes \vert m\rangle\; )= \\
\vspace*{0.1in}
\Psi \biggl(\ \biggl[ \lambda^{2}X_{p}^{+}(x)X_{p}^{-}(x)\ -\ X_{p}^{-}(x)\; \sum_{T_{x}(\overline{\gamma}_{0})}\bold{\hat{A}}\ -\ X_{p}^{+}(x)\; \sum_{\overline{T}_{x}(\overline{\overline{\gamma}}_{0})}\bold{\hat{B}}\\
\vspace*{0.1in}
\\
+\ \sum_{\triangle_{m}\in T_{x}(\overline{\gamma_{0}})}X_{p}^{+}(v_{m})\bold{\hat{B}}\ +\ \sum_{\triangle_{m}\in T_{x}(\overline{\overline{\gamma_{0}})}}X_{p}^{-}(v_{m})\bold{\hat{A}}\ + \frac{\hat{m_{R}}}{\lambda}\ \biggr](\ f^{+}_{\overline{\gamma}_{0}}\otimes\ f^{-}_{\overline{\overline{\gamma}}_{0}})\ \biggr)
\end{array}
\end{equation}
\pagebreak

\section{Expectation value of the physical dilaton operator}
Let $f^{+}_{\overline{s}}\otimes\ f^{-}_{\overline{\overline{s}}}\otimes \vert m>$ be such that $\eta(\ f^{+}_{\overline{s}}\otimes\ f^{-}_{\overline{\overline{s}}}\ \otimes \vert m>\ )\ =\ \Psi$.\\
Whence,\\
\begin{equation}
\langle\ \Psi \vert\ \widehat{y(x)[X^{+}_{p},X^{-}_{p}]}_{can'}\ \vert\Psi\ \rangle\ =\ \Psi(\ \widehat{y(x)[X^{+}_{p},X^{-}_{p}]}^{\dagger}\ f^{+}_{\overline{s_{o}}}\otimes\ f^{-}_{\overline{\overline{s_{0}}}}\ ).
\end{equation}
We make the following choice for triangulation.\\
Given an orbit of diffeomorphism-equivalence class of charge-networks $(\overline{s}\; ,\; \overline{\overline{s}})$ we choose $(\overline{s}_{0}\; ,\; \overline{\overline{s}}_{0})$ such that r($\overline{\overline{\gamma}}_{0}(\overline{\overline{s}}_{0})$) $\subset$ r($\overline{\gamma}_{0}(\overline{s}_{0})$) and over the range r($\overline{\overline{\gamma}}_{0}$), the subgraph of $\overline{\gamma}_{0}$ coincides with $\overline{\overline{\gamma}}_{0}$. (see figure below)
\begin{center}
\includegraphics[scale=0.7]{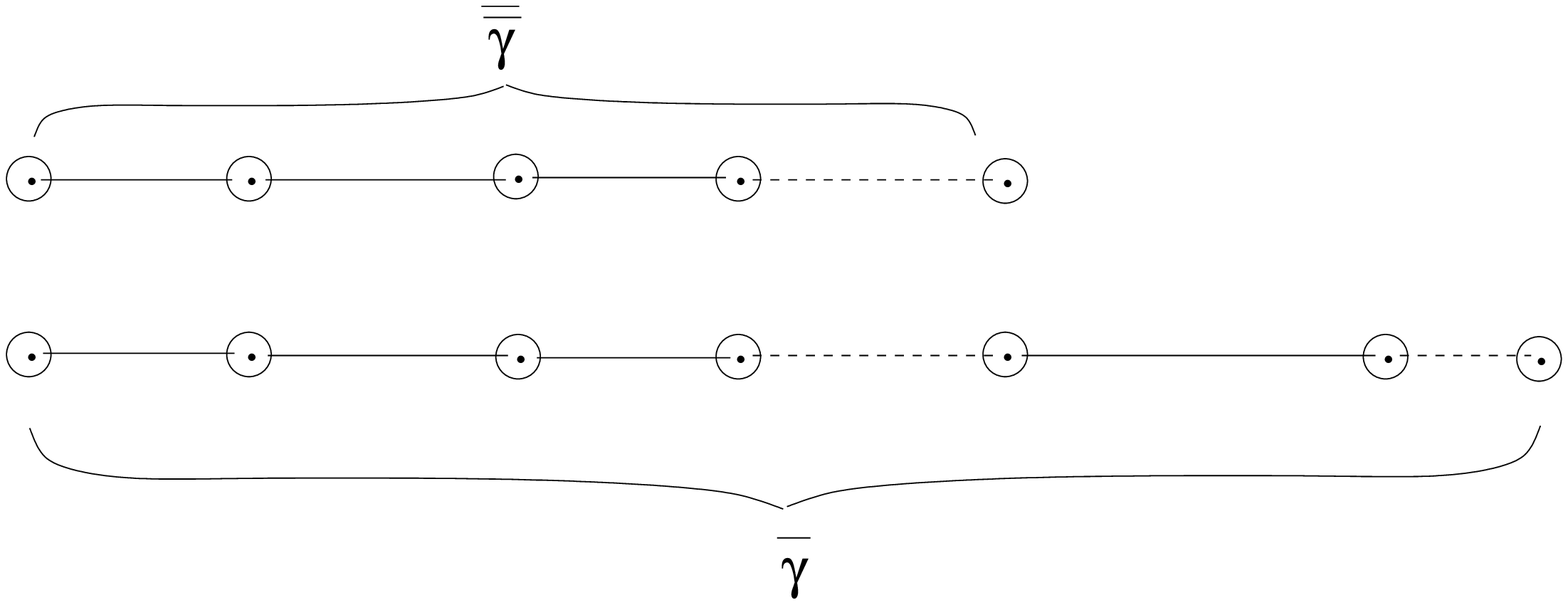}
fig.2
\end{center}
\hspace*{0.2in} We choose the triangulations adapted to $\overline{\gamma}_{0}$ , $\overline{\overline{\gamma}}_{0}$ as $T(\overline{\gamma_{0}})\; =\; T(\overline{\overline{\gamma}}_{0})\; =\; \overline{\gamma}_{0}\; \cup e_{L}\; \cup e_{R}$ where $e_{L}$ and $e_{R}$ are arbitrary 1-simplices from $-\infty$ to initial(left-most) vertex of $\overline{\gamma}_{0}$ and from final(right-most) vertex of $\overline{\gamma}_{0}$ to $\infty$ respectively.\\
Now let us see how each term simplifies.\\
\underline{Term 2}\\
\hspace*{0.2in} It is easy to show that with the above choice of $(\overline{\gamma}_{0}\; ,\; \overline{\overline{\gamma}}_{0})$ , $(T(\overline{\gamma}_{0})\; ,\; T(\overline{\overline{\gamma}}_{0}))$ term-2 can be written as,\\
{\setlength\arraycolsep{1pt}
\begin{equation}
\begin{array}{lll}
\langle \ \Psi \vert\ \frac{X^{-}_{p}(x)}{4}\; \int_{\infty}^{x}d\overline{x}\frac{\widehat{Y_{-}(\overline{x})[X^{+}_{p},X^{-}_{p}]^{2'}}}{X_{p}^{-'}(\overline{x})} \vert\Psi\ \rangle   = &\langle\; \Psi\; [\frac{X^{-}_{p}(x)}{4}\; \int_{\infty}^{x}d\overline{x}\frac{\widehat{Y_{-}(\overline{x})[X^{+}_{p},X^{-}_{p}]^{2}}^{\dagger}}{X_{p}^{-'}(\overline{x})}]\; (f^{+}_{\overline{\gamma}_{0}}\otimes f^{-}_{\overline{\overline{\gamma}}_{0}})&\\
\\
\vspace*{0.1in}
&= \langle\; \Psi[ \frac{X^{-}_{p}(x)}{16\pi} \sum_{\triangle_{m}\in T_{x}}\biggr[\; \frac{(X^{+}_{p}(v_{m})\; -\; X^{+}_{p}(v_{m}))^{2}}{X^{-}_{p}(v_{m+1})\; -\; X^{-}_{p}(v_{m})}&\\
\\
\vspace*{0.1in}
&  \int_{-\infty}^{0}dk d{\overline{k}} \biggr(\; e^{i(k+\overline{k})X_{p}^{+}(v_{m})}(\; -2\sum_{n=0}^{N}e^{-\frac{i}{2}(k+\overline{k})(k^{+}_{n}+k^{+}_{n+1})}\; -\; 4\; )&\\
\\
\vspace*{0.1in}
&  \qquad \quad \quad -e^{-i(k-\overline{k})X_{p}^{+}(v_{m})}(\; -2\sum_{n=0}^{N}e^{-\frac{i}{2}(k-\overline{k})(k^{+}_{n}+k^{+}_{n+1})}\; -\; 4\; )&\\
\\
\vspace*{0.1in}
&  \qquad \quad \quad -e^{i(k-\overline{k})X_{p}^{+}(v_{m})}(\; -2\sum_{n=0}^{N}e^{\frac{i}{2}(k-\overline{k})(k^{+}_{n}+k^{+}_{n+1})}\; -\; 4\; )&\\
\\
\vspace*{0.1in}
&  \qquad \quad \quad +e^{-i(k+\overline{k})X_{p}^{+}(v_{m})}(\; -2\sum_{n=0}^{N}e^{\frac{i}{2}(k+\overline{k})(k^{+}_{n}+k^{+}_{n+1})}\; -\; 4\; )\biggl)\biggl]&\\
\vspace*{0.1in}
& \hspace*{3.0in}  (f^{+}_{\overline{\gamma}_{0}}\otimes f^{-}_{\overline{\overline{\gamma}}_{0}})&
\end{array}
\end{equation}}
where we have set $k^{+}_{0}\; =\; k^{+}_{N+1}\; =\; 0$.
\\
Similarly \underline{Term-3} can be written as,\\
{\setlength\arraycolsep{1pt}
\begin{equation}
\begin{array}{lll}
\langle \ \Psi \vert\ \frac{X^{+}_{p}(x)}{4}\; \int_{-\infty}^{x}d\overline{x}\frac{\widehat{Y^{+}(\overline{x})[X^{+}_{p},X^{-}_{p}]^{2'}}}{X_{p}^{+'}(\overline{x})} \vert\Psi\ \rangle  & = & \langle\; \Psi\; [\frac{X^{+}_{p}(x)}{4}\; \int_{-\infty}^{x}d\overline{x}\frac{\widehat{Y^{+}(\overline{x})[X^{+}_{p},X^{-}_{p}]^{2}}^{\dagger}}{X_{p}^{+'}(\overline{x})}]\; (f^{+}_{\overline{\gamma}_{0}}\otimes f^{-}_{\overline{\overline{\gamma}}_{0}})\\
\\
\vspace*{0.1in}
& = & \langle\; \Psi[ \frac{X^{+}_{p}(x)}{16\pi} \sum_{\triangle_{m}\in \overline{T}_{x}}\biggr[\; \frac{(X^{-}_{p}(v_{m})\; -\; X^{-}_{p}(v_{m}))^{2}}{X^{+}_{p}(v_{m+1})\; -\; X^{+}_{p}(v_{m})}\\
\\
\vspace*{0.1in}
& & \int_{0}^{\infty}dk d{\overline{k}} \biggr(\; e^{i(k+\overline{k})X_{p}^{-}(v_{m})}(\; -2\sum_{n=0}^{M}e^{-\frac{i}{2}(k+\overline{k})(k^{-}_{n}+k^{-}_{n+1})}\; -\; 4\; )\\
\\
\vspace*{0.1in}
& & \qquad \quad \quad -e^{-i(k-\overline{k})X_{p}^{-}(v_{m})}(\; -2\sum_{n=0}^{M}e^{-\frac{i}{2}(k-\overline{k})(k^{-}_{n}+k^{-}_{n+1})}\; -\; 4\; )\\
\\
\vspace*{0.1in}
& & \qquad \quad \quad -e^{i(k-\overline{k})X_{p}^{-}(v_{m})}(\; -2\sum_{n=0}^{M}e^{\frac{i}{2}(k-\overline{k})(k^{-}_{n}+k^{-}_{n+1})}\; -\; 4\; )\\
\\
\vspace*{0.1in}
& & \qquad \quad \quad +e^{-i(k+\overline{k})X_{p}^{-}(v_{m})}(\; -2\sum_{n=0}^{N}e^{\frac{i}{2}(k+\overline{k})(k^{-}_{n}+k^{-}_{n+1})}\; -\; 4\; )\biggl)\biggl]\\
\vspace*{0.1in}
& & \hspace*{3.0in}  (f^{+}_{\overline{\gamma}_{0}}\otimes f^{-}_{\overline{\overline{\gamma}}_{0}})
\end{array}
\end{equation}}
where we have set $k^{-}_{0}\; =\; k^{-}_{M+1}\; =\; 0$.\\
Rest of the terms can be written in a similar fashion and we can write the final expression for $<\widehat{y(x)[X^{+}_{p},X^{-}_{p}]}>$ as follows,\\
\begin{equation}\label{eq:five}
\begin{array}{lll}
\langle \Psi \vert \widehat{y(x)[X^{+}_{p},X^{-}_{p}]}_{can}\vert \Psi \rangle_{phy}\; =\; \\
\vspace*{0.1in}
\Psi \biggr[\; \lambda^{2}X_{p}^{+}(x)X_{p}^{-}(x)\; + \frac{X_{p}^{-}(x)}{16\pi}\sum_{\triangle_{m}\in T_{x}}\mathcal{A}_{m}\; -\; \frac{X_{p}^{+}(x)}{16\pi}\sum_{\triangle_{m}\in T_{x}}\mathcal{B}_{m}\; -\; \frac{1}{16\pi}\sum_{\triangle_{m}\in T_{x}}X^{-}_{p}(v_{m})\mathcal{A}_{m}\\
\vspace*{0.1in}
-\; \frac{1}{16\pi}\sum_{\overline{T}_{x}}X^{+}_{p}(v_{m})\mathcal{B}_{m}\; +\; \frac{\hat{m}_{R}}{\lambda}\; \biggl](f^{+}_{\overline{s_{0}}}\otimes f^{-}_{\overline{\overline{s_{0}}}}\otimes \vert m>)
\end{array}
\end{equation}
Recall that $T_{x}$ is subcomplex from -$\infty$ to x and $\overline{T}_{x}$ is the subcomplex from x to $\infty$.\\
$\mathcal{A}_{m}$ and $\mathcal{B}_{m}$ are respectively given by,\\
\begin{equation}
\begin{array}{lll}
\mathcal{A}_{m} & = & \frac{[X^{+}_{p}(v_{m})\; -\; X^{+}_{p}(v_{m-1})]^{2}}{X_{p}^{-}(v_{m})\; -\; X_{p}^{-}(v_{m-1})}\\
\\
\vspace*{0.2in}
& & \int_{-\infty}^{0}dk d\overline{k} \biggl\lbrace e^{i(k+\overline{k})X_{p}^{+}(v_{m})}(\; -2\sum_{n=0}^{N}e^{-\frac{i}{2}(k+\overline{k})(k^{+}_{n}+k^{+}_{n+1})}\; -\; 4\; )\; +\; c.c.\\
\vspace*{0.1in}
\quad & & - e^{-i(k-\overline{k})X_{p}^{+}(v_{m})}(\; -2\sum_{n=0}^{N}e^{-\frac{i}{2}(k-\overline{k})(k^{+}_{n}+k^{+}_{n+1})}\; -\; 4\; )\; +\; c.c.\biggr\rbrace
\end{array}
\end{equation}
\begin{equation}
\begin{array}{lll}
\mathcal{B}_{m} & = & \frac{[X^{-}_{p}(v_{m})\; -\; X^{-}_{p}(v_{m-1})]^{2}}{X_{p}^{+}(v_{m})\; -\; X_{p}^{+}(v_{m-1})}\\
\\
\vspace*{0.2in}
& & \int_{0}^{\infty}dk d\overline{k} \biggl\lbrace e^{i(k+\overline{k})X_{p}^{-}(v_{m})}(\; -2\sum_{n=0}^{N}e^{-\frac{i}{2}(k+\overline{k})(k^{-}_{n}+k^{-}_{n+1})}\; -\; 4\; )\; +\; c.c.\\
\vspace*{0.1in}
\quad & & - e^{-i(k-\overline{k})X_{p}^{-}(v_{m})}(\; -2\sum_{n=0}^{N}e^{-\frac{i}{2}(k-\overline{k})(k^{-}_{n}+k^{-}_{n+1})}\; -\; 4\; )\; +\; c.c.\biggr\rbrace
\end{array}
\end{equation}
One can further simplify this expression as follows. Using,\\
\begin{equation}
\int_{-\infty}^{0}e^{ikX_{p}^{+}(v_{m})}\; :=\; \int_{-\infty}^{0}e^{ik[X_{p}^{-}(v_{m})\; -\; i\epsilon]}\; =\; \frac{1}{X^{-}_{p}(v_{m})\; -\; i\epsilon}\; =\; P(\frac{1}{X_{p}^{-}(v_{m})})\; -\; i\epsilon \delta(X_{p}^{-}(v_{m}))
\end{equation}
one can show that $\mathcal{A}_{m}$, $\mathcal{B}_{m}$  simplify to,\\
\begin{equation}
\begin{array}{lll}
\mathcal{A}_{m} & = & \frac{[X^{+}_{p}(v_{m})\; -\; X^{+}_{p}(v_{m-1})]^{2}}{X_{p}^{-}(v_{m})\; -\; X_{p}^{-}(v_{m-1})}\\
\\
\vspace*{0.2in}
& & \biggl\lbrace -4\sum_{n=0}^{N+1} P(\frac{1}{X^{+}_{p}(v_{m})\; -\; \frac{1}{2}(k^{+}_{n}+k^{+}_{n+1})})^{2}\; -\; 8\; P(\frac{1}{X_{p}^{+}(v_{m})})^{2}\biggr\rbrace
\end{array}
\end{equation}
\begin{equation}
\begin{array}{lll}
\mathcal{B}_{m} & = & \frac{[X^{-}_{p}(v_{m})\; -\; X^{-}_{p}(v_{m-1})]^{2}}{X_{p}^{+}(v_{m})\; -\; X_{p}^{+}(v_{m-1})}\\
\\
\vspace*{0.2in}
& & \biggl\lbrace -4\sum_{n=0}^{N+1} P(\frac{1}{X^{-}_{p}(v_{m})\; -\; \frac{1}{2}(k^{-}_{n}+k^{-}_{n+1})})^{2}\; -\; 8\; P(\frac{1}{X_{p}^{-}(v_{m})})^{2}\biggr\rbrace
\end{array}
\end{equation}


\begin{thebibliography}{10}

\bibitem{moore}
 P.~ Ginsparg, G.~ Moore,
\newblock  Lectures on 2D gravity and 2D string theory(TASI 1992).
\newblock{hep-th/9304011 }.

\bibitem{cghs}
C.G.~ Callan, S.B.~ Giddings, J.A.~ Harvey, and A.~ Strominger,
\newblock Evanescent Black Holes.
\newblock {Phys.Rev. D45 (1992) 1005-1009}.

\bibitem{rst}
 J.~ Russo, L.~ Susskind, L.~ Thorlacius
\newblock Black Hole Evaporation in 1+1 Dimensions.
\newblock {Phys.Lett. B292 (1992) 13-18}.

\bibitem{s}
A. ~ Strominger,
\newblock Fadeev-Popov Ghosts and 1+1 Dimensional Black Hole Evaporation.
\newblock{Phys.Rev. D46 (1992) 4396-4401}.

\bibitem{g-s}
S.B.~ Giddings, A.~ Strominger
\newblock Quantum Theories of Dilaton Gravity
\newblock {Phys.Rev. D47 (1993) 2454-2460}.

\bibitem{krv}
 K.~ V.~ Kuchar, J.~ D.~ Romano, M.~ Varadarajan,
\newblock Dirac Constraint Quantization of a Dilatonic Model of Gravitational Collapse
\newblock { Phys.Rev. D55 (1997) 795-808}.

\bibitem{jackiw1}
 E.~ Benedict, R.~ Jackiw, H.~ J.~ Lee,
\newblock Functional Schroedinger and BRST Quantization of (1+1)--Dimensional Gravity.
\newblock{Phys.Rev. D54 (1996) 6213-6225}.

\bibitem{jackiw2}
 D.~ Cangemi, R.~ Jackiw, B.~ Zwiebach,
\newblock Physical States in Matter-Coupled Dilaton Gravity
\newblock {Annals Phys. 245 (1996) 408-444}.

\bibitem{mikovic}
A.~ Mikovic,
\newblock Unitary Theory of Evaporating 2D Black Holes.
\newblock{CQG 13 (1996) 209-220}.

\bibitem{th0}
  T.~ Thiemann,
\newblock     Introduction to Modern Canonical Quantum General Relativity.
\newblock{gr-qc/0110034}.

\bibitem{r}
  C.~ Rovelli,
\newblock      Quantum gravity.
\newblock{Cambridge, UK: Univ. Pr. (2004) 455 p}.

\bibitem{ashtekar1}
A.~ Ashtekar, J.~ Lewandowski, H.~ Sahlmann,
\newblock  Polymer and Fock representations for a Scalar field.
\newblock{Class.Quant.Grav. 20 (2003) L11-1}.

\bibitem{ashtekar2}
 A.~ Ashtekar, J.~ Lewandowski,
\newblock   Relation between polymer and Fock excitations.
\newblock{Class.Quant.Grav. 18 (2001) L117-L128}.

\bibitem{hajicek1}
  P.~ Hajicek, C.~ J.~ Isham,
\newblock   Perennials and the Group-Theoretical Quantization of a Parametrized Scalar Field on a Curved Background.
\newblock{J.Math.Phys. 37 (1996) 3522-3538}.

\bibitem{hajicek2}
  P.~ Hajicek,
\newblock   Time evolution and observables in constrained systems.
\newblock{Class.Quant.Grav. 13 (1996) 1353-1376}.

\bibitem{hajicek3}
  P.~ Hajicek,
\newblock    Time evolution of observable properties of parametrized systems.
\newblock{Nucl.Phys.Proc.Suppl. 57 (1997) 115-124}.

\bibitem{bianca}
 B.~ Dittrich,
\newblock   Partial and Complete Observables for Hamiltonian Constrained Systems.
\newblock{gr-qc/0411013}.

\bibitem{giddings}
  S.~ Giddings, W.~ Nelson
\newblock   Quantum Emission from Two-dimensional Black Holes.
\newblock{Phys.Rev. D46 (1992) 2486-2496}.

\bibitem{hajicek4}
 P.~ Hajicek,  J.~ Kijowski
\newblock    Covariant gauge fixing and Kuchar decomposition.
\newblock{Phys.Rev. D61 (2000) 024037}.

\bibitem{m1}
 M.~ Varadarajan
\newblock    Classical and quantum geometrodynamics of 2d vacuum dilatonic black holes.
\newblock{Phys.Rev. D52 (1995) 7080-7088}.

\bibitem{k1}
 K.~ V.~ Kuchar,
\newblock    Geometrodynamics of Schwarzschild Black Holes.
\newblock{Phys.Rev. D50 (1994) 3961-3981}.

\bibitem{m2}
 M.~ Varadarajan
\newblock    Quantum gravity effects in the CGHS model of collapse to a black hole.
\newblock{Phys.Rev. D57 (1998) 3463-3473}.

\bibitem{th1}
 H.~ Sahlmann, T.~ Thiemann,
\newblock     On the superselection theory of the Weyl algebra for diffeomorphism invariant quantum gauge theories.
\newblock{gr-qc/0302090}.

\bibitem{th2}
  J.~ Lewandowski, A.~ Okolow, H.~ Sahlmann, T.~ Thiemann,
\newblock     Uniqueness of diffeomorphism invariant states on holonomy-flux algebras.
\newblock{gr-qc/0504147}.

\bibitem{th3}
  T.~ Thiemann,
\newblock     The LQG -- String: Loop Quantum Gravity Quantization of String Theory I. Flat Target Space.
\newblock{Class.Quant.Grav. 23 (2006) 1923-1970}.

\bibitem{marolf}
 D.~ Giulini, D.~ Marolf
\newblock    On the Generality of Refined Algebraic Quantization.
\newblock{Class.Quant.Grav. 16 (1999) 2479-2488}.

\bibitem{ashtekar3}
 A.~ Ashtekar, J.~ Lewandowski, D.~ Marolf, J.~ Mourao, T.~ Thiemann
\newblock    Quantization of diffeomorphism invariant theories of connections with local degrees of freedom.
\newblock{J.Math.Phys. 36 (1995) 6456-6493}.

\bibitem{rovelli}
 C.~ Rovelli
\newblock      What Is Observable In Classical And Quantum Gravity?.
\newblock{Class.Quant.Grav.8:297-316,1991}.



\bibitem{torre}
 C.~ G.~ Torre
\newblock     A Complete set of observables for cylindrically symmetric gravitational fields.
\newblock{Class.Quant.Grav.8:1895-1912,1991}.

\bibitem{urs}
 U.~ Schreiber
\newblock     DDF and Pohlmeyer invariants of (super)string.
\newblock{JHEP 0405 (2004) 027}.

\bibitem{th4}
  H.~ Sahlmann, T.~ Thiemann, O.~ Winkler,
\newblock      Coherent States for Canonical Quantum General Relativity and the Infinite Tensor Product Extension.
\newblock{Nucl.Phys. B606 (2001) 401-440}.




\end{thebibliography}
\end{document}